\documentclass[amsmath,reprint,prb,longbibliography,superscriptaddress,bibnotes]{revtex4-2}
\usepackage{amsthm,amssymb,amsfonts,graphicx,verbatim, xcolor,bm} 
\usepackage{hyperref} 
\usepackage[utf8]{inputenc} 
\usepackage{siunitx} 
\usepackage{braket}
\usepackage{ulem}
\graphicspath{{figures}}

\begin{document}

\title{Sublattice structure and topology in spontaneously crystallized electronic states}
\author{Yongxin Zeng}
\email{yz4788@columbia.edu}
\affiliation{Department of Physics, Columbia University, New York, NY 10027}
\author{Daniele Guerci}
\affiliation{Center for Computational Quantum Physics, Flatiron Institute, New York, NY 10010}
\author{Valentin Cr\'epel}
\affiliation{Center for Computational Quantum Physics, Flatiron Institute, New York, NY 10010}
\author{Andrew J. Millis}
\affiliation{Department of Physics, Columbia University, New York, NY 10027}
\affiliation{Center for Computational Quantum Physics, Flatiron Institute, New York, NY 10010}
\author{Jennifer Cano}
\affiliation{Department of Physics and Astronomy, Stony Brook University, Stony Brook, NY 11794}
\affiliation{Center for Computational Quantum Physics, Flatiron Institute, New York, NY 10010}

\begin{abstract}
The prediction and realization of the quantum anomalous Hall effect are often intimately connected to honeycomb lattices in which the sublattice degree of freedom plays a central role in the nontrivial topology. Two-dimensional Wigner crystals, on the other hand, form triangular lattices without sublattice degrees of freedom, resulting in a topologically trivial state. In this Letter, we discuss the possibility of spontaneously formed honeycomb-lattice crystals that exhibit the quantum anomalous Hall effect. Starting from a single-band system with nontrivial quantum geometry, we derive the mean-field energy functional of a class of crystal states and express it as a model of sublattice pseudospins in momentum space. We find that nontrivial quantum geometry leads to extra terms in the pseudospin model that break an effective `time-reversal symmetry' and favor a topologically nontrivial pseudospin texture. When the effects of these extra terms dominate over the ferromagnetic exchange coupling between pseudospins, the anomalous Hall crystal state becomes energetically favorable over the trivial Wigner crystal state.
\end{abstract}

\maketitle


{\it Introduction.---}
Topology and spontaneously broken symmetry are two central themes of modern condensed matter physics. One prototypical broken symmetry phase is the Wigner crystal (WC) \cite{wigner1934}, a phase characterized by spontaneously broken translational invariance and trivial topology. In two dimensions (2D), semiclassical calculations in the low-density limit predict \cite{bonsall1977static} that the lowest energy configuration of electrons moving in a uniform positive background and interacting via the conventional Coulomb interaction is a triangular-lattice Wigner crystal. Theoretical and experimental studies \cite{tanatar1989ground, drummond2009phase, spivak2004phases, zarenia2017wigner, goldman1990evidence, yoon1999wigner, hossain2020observation, zhou2021bilayer, smolenski2021signatures, li2021imaging, regan2020mott, xiang2024quantum} over the last few decades have established Wigner crystals as one of the prototypical broken symmetry states of strongly interacting electron systems.

The quantum anomalous Hall (QAH) insulator \cite{chang2023colloquium} is a topologically nontrivial insulating state that exhibits a quantized Hall conductance in the absence of an external magnetic field. First predicted by Haldane in a honeycomb-lattice model \cite{haldane1988model}, it has been realized experimentally in magnetic topological insulators \cite{chang2013experimental, chang2015high, deng2020quantum} and more recently in moir\'e superlattices \cite{li2021quantum, tao2024valley, serlin2020intrinsic, han2023correlated, lu2023fractional, han2023large} where the low-energy QAH physics can often be understood via a mapping to the Haldane model in a honeycomb superlattice \cite{wu2019topological,devakul2021magic,crepel2023anomalous}. 

The sublattice degree of freedom is crucial for the nontrivial topology of QAH insulators in honeycomb lattices. In fact, it was shown in recent work \cite{zeng2024gate,tan2024designing, su2022massive,crepel2023chiraljen} that when a honeycomb-superlattice modulation is applied to a gapped 2D system with nontrivial band geometry, the lowest miniband is topologically nontrivial under mild assumptions. In these cases, translation invariance is explicitly broken. The question of the circumstances under which a {\it spontaneously broken} translation invariance may lead to a topologically nontrivial QAH state remains open.

The coexistence of spontaneously broken translation invariance and conventional (magnetic field induced) quantum Hall effects was discussed in pioneering papers by Halperin, Te\v{s}anovi\'c, and Axel \cite{halperin1986compatibility, tesanovic1989hall}. This work was a proof of principle, based on a model with specifically tuned interactions. The possibility of a spontaneously formed crystal phase that exhibits the QAH effect in zero applied field, however, has apparently not been considered until very recently. Following the discovery of integer and fractional QAH effects in rhombohedral pentalayer graphene aligned to a hexagonal boron nitride substrate \cite{han2023correlated, lu2023fractional, han2023large}, mean-field calculations found \cite{dong2023anomalous, dong2023theory, zhou2023fractional, guo2023theory, kwan2023moire} a robust QAH insulator state at filling factor $\nu=1$. The experimental devices involved a moir\'e potential that explicitly broke the translational invariance and the theoretical studies therefore were based on models that also explicitly broke translation invariance.  Surprisingly, the QAH insulator state was found theoretically to persist in the limit of vanishing moir\'e potential, implying that a spontaneous translational symmetry breaking could lead to an `anomalous Hall crystal (AHC)' state.

\begin{figure}
     \includegraphics[width=0.9\linewidth]{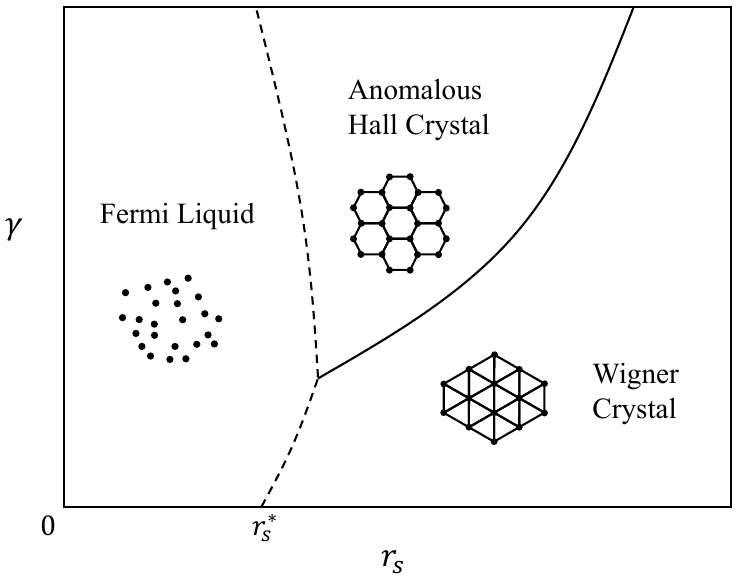}
     \caption{Schematic phase diagram in the plane of interaction strength $r_s$ and Berry curvature concentration $\gamma$ showing regions of Fermi liquid, conventional Wigner crystal, and anomalous Hall crystal phases. The precise definitions of $r_s$ and $\gamma$ are provided in a later part of the paper. $r_s^*$ is the critical interaction strength for the Wigner crystallization phase transition in systems with trivial quantum geometry. 
     Our theory focuses on the transition between two crystal phases (solid line); melting of crystals is not explicitly considered in our theory and the dashed line is speculation based on the Lindemann criterion.}
     \label{fig:schematic_phase}
\end{figure}

In this Letter we investigate theoretically the circumstances under which a topologically nontrivial electron crystal may occur. We use a simple yet general model based a single band of electrons that we view as representing the low-energy physics of a multiband system. The quantum geometry appears in the Hamiltonian as form factors in the projection of Coulomb interactions onto the low-energy band. We use this Hamiltonian to study the energetic competition between WC and AHC states. Crucial to our analysis is the observation that the WC and AHC states have the same Bravais lattice but different sublattice structures. This enables the construction of a sublattice pseudospin representation of the physics that provides an interpolation between the topologically trivial WC and nontrivial AHC states. Combining analytic derivation of a sublattice pseudospin model and numerical self-consistent mean-field calculations, we find that AHCs are stabilized by strong enough Berry curvature concentration at intermediate interaction strengths, implying the phase diagram shown in Figure ~\ref{fig:schematic_phase}.

{\it Model.---}
We consider a 2D electron system described by a single band with arbitrary dispersion consistent with $C_6$ symmetry and nontrivial quantum geometry described by the Hamiltonian $H = H_{\rm kin} + H_{\rm int}$. The kinetic energy is
\begin{equation}
H_{\rm kin} = \sum_{\bm p} E_{\bm p} c_{\bm p}^{\dagger} c_{\bm p},
\label{eq:Hkin}
\end{equation}
where $E_{\bm p}$ is the band dispersion and $c_{\bm p}$ ($c_{\bm p}^{\dagger}$) is the annihilation (creation) operator of electrons with momentum $\bm p$. We view $E_{\bm p}$ as the lowest-lying band of a multiband system and the momentum $\bm p$ runs over the large Brillouin zone of the microscopic lattice describing this multiband system. Because the physics of interest involves a low density of electrons in this band, leading to a long-period superlattice, we may treat the large Brillouin zone as an infinite 2D momentum space. Interactions between electrons are described by the Hamiltonian
\begin{equation}
H_{\rm int} = \frac{1}{2\mathcal{A}} \sum_{\bm p \bm p' \bm q} V_{\bm q} \Lambda_{\bm p + \bm q, \bm p} \Lambda_{\bm p', \bm p' + \bm q} c_{\bm p + \bm q}^{\dagger} c_{\bm p'}^{\dagger} c_{\bm p' + \bm q} c_{\bm p},
\end{equation}
where $\mathcal{A}$ is the area of the 2D system. Anticipating the honeycomb-lattice structure of spontaneously formed crystals, we require that both $E_{\bm p}$ and $V_{\bm q}$ preserve $C_6$ rotational symmetry. The quantum geometry is encoded in the form factor $\Lambda_{\bm p', \bm p} = \braket{u_{\bm p'}|u_{\bm p}}$, where $\ket{u_{\bm p}}$ is the periodic part of the Bloch wavefunction of the projected band. $C_6$ symmetry imposes the constraint $\Lambda_{C_6 \bm p', C_6 \bm p} = \Lambda_{\bm p', \bm p}$. For a trivial band with vanishing Berry curvature, $\Lambda_{\bm p', \bm p} \in \mathbb{R}$ by a proper gauge choice. This implies an emergent $C_2 \mathcal{T}$ symmetry where $\mathcal{T}$ is an effective `time-reversal' operator \footnote{Note that $\mathcal{T}$ is not the physical time-reversal operator, but rather an anti-unitary operator that acts like time-reversal within a single valley. The complete physical system contains another valley that is the physical time-reversal partner of the band we describe and transforms differently under the effective time-reversal $\mathcal{T}$, but we assume it is at higher energy due to either explicit or spontaneous breaking of the physical time-reversal symmetry.}. For a generic band with nontrivial quantum geometry, $\Lambda_{\bm p', \bm p}$ is in general complex and the effective time-reversal symmetry (TRS) \footnote{Throughout this paper, by TRS we always refer to the effective time-reversal symmetry that acts within a single valley.} is broken. 

The Hartree-Fock potential defines a Bravais lattice common to both the WC and AHC states, which are distinguished by different sublattice structures. We may view the triangular lattice WC as the state with one sublattice of the AHC honeycomb occupied and the other one empty. We are interested in the lowest-lying bands of the long-period superlattice; these are defined in terms of the original microscopic states via the sublattice basis $a_{\bm k}^{\dagger} = \sum_{\bm g} A_{\bm k + \bm g} c_{\bm k + \bm g}^{\dagger}$ and $b_{\bm k}^{\dagger} = \sum_{\bm g} B_{\bm k + \bm g} c_{\bm k + \bm g}^{\dagger}$, where $A_{\bm p}, B_{\bm p}$ are momentum-space wavefunctions of the localized sublattice orbitals and $\bm g$ sums over reciprocal lattice vectors of the long-period superlattice whose lattice constant is determined by the electron density. To comply with the point-group symmetries of the honeycomb lattice, each sublattice basis state has TRS and $C_3$ rotational symmetry around its center, and the two sublattices are related by $C_2$ rotation around the hexagon center (see Supplemental Material for details). A general Hartree-Fock ground state is a superposition of two sublattice basis states:
\begin{equation} \label{eq:Psi}
\ket{\Psi} = \prod_{\bm k \in {\rm mBZ}} \left(\cos\frac{\theta_{\bm k}}{2} a_{\bm k}^{\dagger} + e^{i\phi_{\bm k}} \sin\frac{\theta_{\bm k}}{2} b_{\bm k}^{\dagger} \right) \ket{0}.
\end{equation}
Here $\ket{0}$ is the vacuum state and $\bm k$ runs over the mini Brillouin zone (mBZ) of the long-period superlattice.  The polar and azimuthal angles $(\theta_{\bm k}, \phi_{\bm k})$ define a {\it sublattice pseudospin} at each momentum site in the mBZ that can be alternatively represented by a unit vector
\begin{equation} \label{eq:n_k}
\bm{n_k} = (\sin\theta_{\bm k} \cos\phi_{\bm k}, \sin\theta_{\bm k} \sin\phi_{\bm k}, \cos\theta_{\bm k}).
\end{equation}
In the language of sublattice pseudospins, triangular-lattice WCs correspond to out-of-plane polarized states (e.g., $\theta_{\bm k} \approx 0$) and honeycomb-lattice AHCs are states in which the pseudospins form a skyrmion texture in the mBZ and the net out-of-plane polarization vanishes. The pseudospin texture and the precise forms of the sublattice basis states are obtained by variational minimization of energy $\braket{\Psi | H | \Psi}$. The Berry curvature of the ground state $\ket{\Psi}$ is given by the winding of $\bm n_{\bm k}$:
\begin{equation}
    \Omega_{\bm k}=\frac{1}{2}{\bm n}_{\bm k}\cdot(\partial_{k_x}{\bm n}_{\bm k}\times\partial_{k_y}{\bm n}_{\bm k}),
\end{equation}
and the Chern number is $C=\int_{\rm mBZ} d^2{\bm k}\Omega_{\bm k}/(2\pi)$. The WC state has $C=0$ and the AHC state has $C=\pm 1$ depending on the pseudospin texture.

{\it Pseudospin order.---}
To compare the energy of states with different sublattice pseudospin order, we calculate the energy expectation value of a generic state of the form given in Eq.~\eqref{eq:Psi}. Up to a constant energy independent of pseudospin texture, the mean-field energy functional takes the form \cite{macdonald2012pseudospin, min2008pseudospin}
\begin{equation} \label{eq:E_MF}
\begin{split}
&E_{\rm MF}[\theta, \phi] \equiv \braket{\Psi | H | \Psi} \\
&= -\sum_{\bm k} \bm{h_k} \cdot \bm{n_k} - \frac{1}{2} \sum_{\alpha\beta} \sum_{\bm k \bm k'} J_{\bm k \bm k'}^{\alpha\beta} n_{\bm k}^{\alpha} n_{\bm k'}^{\beta},
\end{split}
\end{equation}
where the $\alpha, \beta$ indices run over Cartesian coordinates $x,y,z$. The problem is thus transformed into an effective spin model in momentum space: $\bm{h_k}$ is an effective Zeeman field that acts on sublattice pseudospins, and $J_{\bm k \bm k'}^{\alpha\beta} = J_{\bm k' \bm k}^{\beta\alpha}$ are coupling coefficients between pseudospins at different momentum sites. While the full expressions of $\bm{h_k}$ and $J_{\bm k \bm k'}^{\alpha\beta}$ are lengthy (see Supplemental Material), the symmetry analysis and physical arguments given below make their qualitative features clear.

\begin{figure*}
    \centering
    \includegraphics[width=0.29\linewidth]{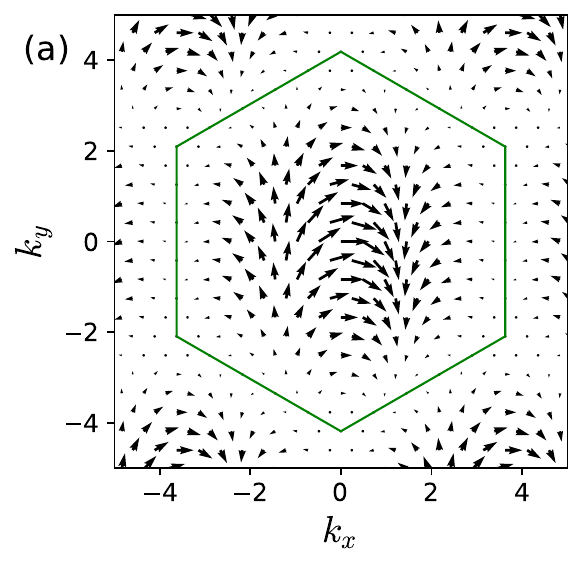}
    \includegraphics[width=0.29\linewidth]{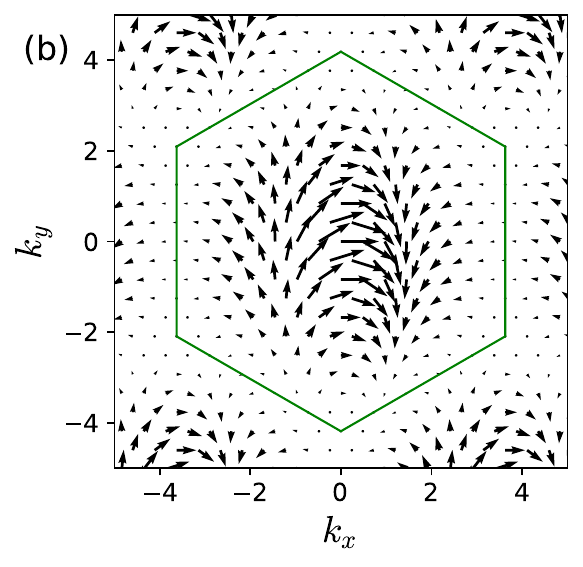}
    \includegraphics[width=0.376\linewidth]{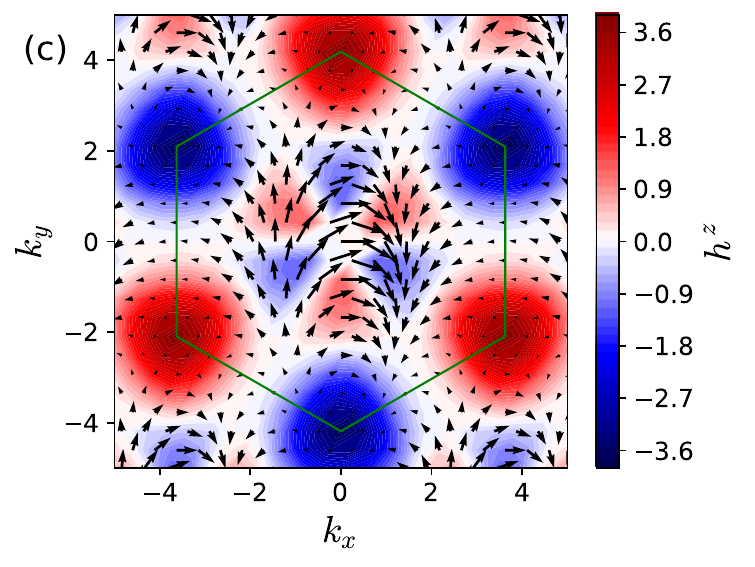}
    \caption{Pseudospin Zeeman field $\bm{h_k}$ in units of the kinetic energy scale $\mathcal{E}_{\rm kin}$. The arrows represent the in-plane components $(h_{\bm k}^x, h_{\bm k}^y)$, scaled down by a factor of 40 and plotted on the axis scale, and the colors represent the out-of-plane component $h_{\bm k}^z$. The green hexagon shows the mBZ boundary. (a) shows the kinetic energy part and (b)-(c) show the Hartree-Fock part at $\gamma=0$ and $\gamma=0.4$ respectively. Other parameters used in the calculations include localization length $l=0.25$, interaction strength $r_s=20$, and winding number $N=3$.}
    \label{fig:h_k}
\end{figure*}

The pseudospin Zeeman field has contributions from both the bare kinetic energy and the interactions; physically, the interaction contribution to $\bm{h_k}$ arises from the Hartree-Fock potential produced by the average electron distribution. Because the average charge density is honeycomb-shaped, the mean-field potential is $C_6$-symmetric. Analogous to graphene, when TRS is preserved, all $\bm{h_k}$ are in plane ($h_{\bm k}^z = 0$) and form vortices of opposite chirality around the Dirac points $K$ and $K'$. See Supplemental Material for a proof by symmetry analysis and Fig.~\ref{fig:h_k}(a) for graphical representation. 

In addition to pseudospin Zeeman fields, interactions also give rise to coupling between pseudospins. In the limit of small $|\bm k - \bm k'| \ll a^{-1}$ where $a$ is the superlattice constant, the dominant coupling coefficients are
\begin{equation}
J_{\bm k \bm k'}^{xx} \approx J_{\bm k \bm k'}^{yy} \approx J_{\bm k \bm k'}^{zz} \approx \frac{1}{2A} V_{\bm k' - \bm k},
\end{equation}
which represent Heisenberg ferromagnetic coupling between pseudospins. Next-order expansion suggests that out-of-plane coupling $J^{zz}$ is slightly stronger than in-plane coupling $J^{xx}, J^{yy}$. Since the pseudospin texture that follows $\bm{h_k}$ (i.e., $\bm{n_k} = \bm{h_k}/|\bm{h_k}|$) is singular around the Dirac points and leads to high exchange energy cost, spontaneous breaking of sublattice symmetry occurs when interactions get strong. Thus, in the strong-interaction limit (large-$r_s$ limit in Fig.~\ref{fig:schematic_phase}), all electrons are polarized to one of the sublattices, forming a triangular-lattice WC.

A similar mean-field theory study has been carried out in the context of graphene \cite{macdonald2012pseudospin, min2008pseudospin}, where an explicit translational symmetry breaking occurs due to the periodic lattice potential of graphene. Our case differs from these previous works in two important ways. Due to the absence of explicit translational symmetry breaking, the graphene-like state in the weak-interaction limit where kinetic energy dominates is likely an artifact of the sublattice basis construction and the true ground state is a Fermi liquid. More importantly, in the presence of nontrivial quantum geometry, we find that spontaneous breaking of translational symmetry leads to a topologically nontrivial honeycomb-lattice state -- the AHC state -- in a range of intermediate interaction strength.

{\it Effects of quantum geometry.---}
TRS combined with $C_2$ rotational symmetry guarantees invariance of energy with respect to sublattice inversion $n_{\bm k}^z \to -n_{\bm k}^z$ and therefore vanishing of $h^z$, $J^{zx}$, and $J^{zy}$. When TRS is broken by nontrivial form factors, all coefficients are generically nonzero. A nonzero $h^z$ explicitly gaps out the Dirac cones at $K$ and $K'$. In addition, $J^{zx}, J^{zy}$ couplings together with the in-plane components of pseudospins $n^x, n^y$ effectively generate another out-of-plane Zeeman field. $C_2$ symmetry implies that the $z$-component of the net effective Zeeman field is opposite at two Dirac points. The pseudospin texture that is aligned with the effective Zeeman field thus forms a topologically nontrivial skyrmion texture in the mBZ. If $n_K^z = +1$ and $n_{K'}^z = -1$, the Chern number is $C=+1$; the opposite case $n_K^z = -1, n_{K'}^z = +1$ leads to $C=-1$. The locally smooth pseudospin texture also implies lower exchange energy cost and higher stability against sublattice polarized states. Overall, TRS breaking makes the topologically nontrivial AHC state more favorable.

To make the analysis quantitative, we now provide an explicit example where the AHC is stabilized for sufficient TRS breaking. We assume quadratic dispersion $E_{\bm p} = \hbar^2 p^2/2m$ and gate-screened Coulomb interaction $V_{\bm q} = (2\pi e^2/q) \tanh(qd/2)$ where $d$ is the distance to gate. In terms of lattice constant $a$, the kinetic energy scale is $\mathcal{E}_{\rm kin} = \hbar^2/2ma^2$ and the interaction energy scale is $\mathcal{E}_{\rm int} = e^2/a$. Below we express all lengths in units of $a$ and energies in units of $\mathcal{E}_{\rm kin}$, and introduce the $r_s$ parameter by $\pi(r_s \hbar^2/me^2)^2 = \sqrt{3}a^2/2$. In this language $\mathcal{E}_{\rm int} = r_s \sqrt{8\pi/\sqrt{3}}$. Since $d$ is a constant length independent of $a$, $d/a \propto 1/r_s$. In our calculations we take $d/a=50/r_s$.

The sublattice basis states in the variational wavefunction \eqref{eq:Psi} are constructed by solving the problem of an electron moving in a honeycomb-lattice potential in the plane-wave basis and Wannierizing the lowest two bands (see Supplemental Material for details). The potential strength controls the localization length $l$ of the constructed basis states.

To model a band with nontrivial quantum geometry, we take the Bloch wavefunction $\ket{u_{\bm p}}$ as a two-component spinor in the basis of internal microscopic orbitals. Partly motivated by recent work on rhombohedral multilayer graphene \cite{dong2023anomalous, dong2023theory, zhou2023fractional, guo2023theory, kwan2023moire}, we take $\ket{u_{\bm p}} = (\cos(\alpha_{\bm p}/2), e^{i\beta_{\bm p}} \sin(\alpha_{\bm p}/2))$ where $\alpha_{\bm p} = \arctan(\gamma |\bm p|)$ and $\beta_{\bm p} = N \arg(p_x+ip_y)$. Here $N \in \mathbb{Z}_+$ is the winding number and emulates the number of graphene layers, $\gamma \geq 0$ describes the concentration of Berry curvature near the origin. Since momentum is measured in units of the long-period superlattice constant, when $\gamma\sim 1$ Berry curvature is concentrated on the scale of the mBZ. We use $N=3$ for our calculations below unless otherwise stated, although qualitatively similar results are also observed for other winding numbers as shown in the Supplemental Material.

\begin{figure*}
    \centering
    \includegraphics[width=0.29\linewidth]{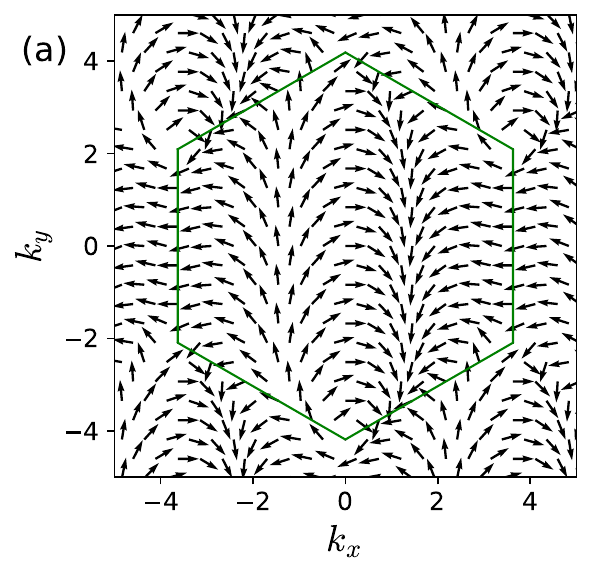}
    \includegraphics[width=0.375\linewidth]{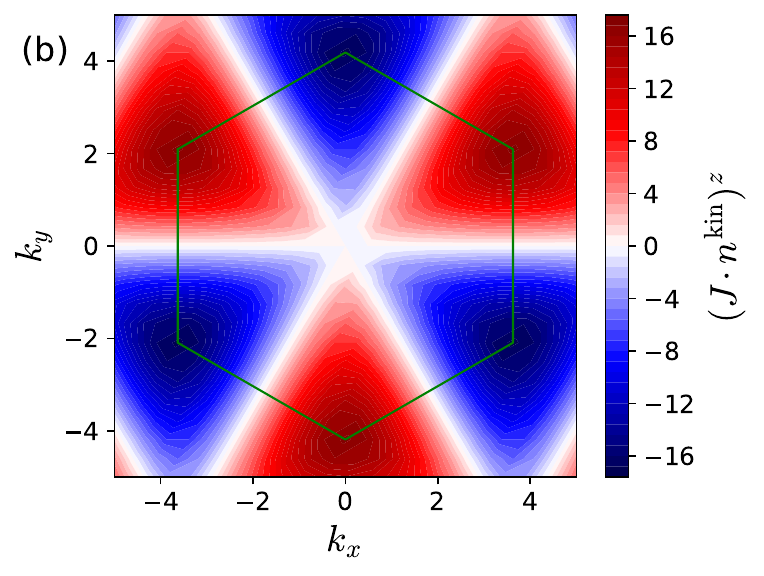}
    \includegraphics[width=0.305\linewidth]{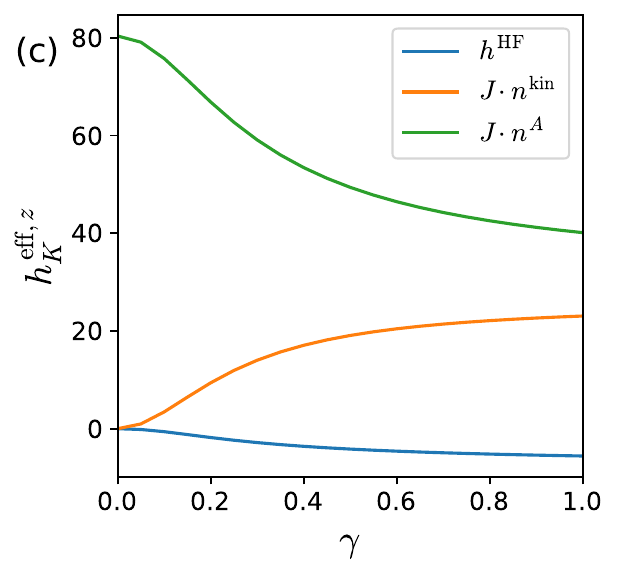}
    \caption{(a) Pseudospin texture that follows the kinetic energy part of Zeeman field $\bm{n_k}^{\rm kin} = \bm{h_k}^{\rm kin} / |\bm{h_k}^{\rm kin}|$. (b) Effective out-of-plane Zeeman field generated by the pseudospins in (a) at $\gamma=0.4$. (c) Effective out-of-plane Zeeman field at $K$ point as a function of $\gamma$. The blue curve represents the Hartree-Fock part of Zeeman field $h_K^{{\rm HF},z}$, and the orange and green curves respectively represent the effective field generated by $\bm{n_k}^{\rm kin}$ and the one generated by the sublattice polarized state $\bm{n_k}^A = (0,0,1)$. Other parameters used in the calculations include $l=0.25$, $r_s=20$, and $N=3$.}
    \label{fig:hz_Jn}
\end{figure*}

Fig.~\ref{fig:h_k} shows the pseudospin Zeeman fields $\bm{h_k}$ computed with a pair of sublattice basis states with localization length $l=0.25$. The kinetic energy part $\bm{h_k}^{\rm kin}$ (Fig.~\ref{fig:h_k}(a)) is independent of form factors. $C_6$ symmetry dictates that it is all in plane and forms vortices of opposite chirality around $K = (2\pi/\sqrt{3}, 2\pi/3)$ and $K' = -K$. When the form factor is trivial ($\gamma=0$), the Hartree-Fock part $\bm{h_k}^{\rm HF}$ (Fig.~\ref{fig:h_k}(b)) forms a similar in-plane pattern as the kinetic part. As $\gamma$ increases and breaks TRS, $\bm{h_k}^{\rm HF}$ gains an out-of-plane component that is opposite at $K$ and $K'$. See Fig.~\ref{fig:h_k}(c) for the case of $\gamma=0.4$.

In the mean-field picture, coupling between pseudospins generates an effective Zeeman field $h_{\bm k}^{{\rm eff},\alpha} = \sum_{\beta, \bm k'} J_{\bm k \bm k'}^{\alpha\beta} n_{\bm k'}^{\beta}$ that depends on the pseudospin texture. To see how $J^{zx}$ and $J^{zy}$ couplings lead to a topologically nontrivial state, we consider the in-plane pseudospin texture that follows the kinetic Zeeman field $\bm{n_k}^{\rm kin} = \bm{h_k}^{\rm kin} / |\bm{h_k}^{\rm kin}|$ (Fig.~\ref{fig:hz_Jn}(a)) and calculate the out-of-plane Zeeman field it generates in the mBZ:
\begin{equation}
h_{\bm k}^{{\rm eff}, z} = \sum_{\bm k'} (J_{\bm k \bm k'}^{zx} n_{\bm k'}^{{\rm kin},x} + J_{\bm k \bm k'}^{zy} n_{\bm k'}^{{\rm kin},y}) \equiv (J\cdot n^{\rm kin})_{\bm k}^z.
\end{equation}
As shown in Fig.~\ref{fig:hz_Jn}(b), the effective field is an odd function of $\bm k$. Comparison between Fig.~\ref{fig:h_k}(c) and Fig.~\ref{fig:hz_Jn}(b) shows that $\bm{h}^{\rm HF}$ and $J\cdot n^{\rm kin}$ have opposite effects on the sublattice polarization at $K$ and $K'$ points and thus push towards opposite AHC states \footnote{Note that this is not a general result for any form factors. For example, for $N=1$ both $\bm h^{\rm HF}$ and $J\cdot n^{\rm kin}$ favor $C=+1$. See Supplemental Material for results at other winding numbers.}; $\bm h^{\rm HF}$ alone leads to a $C=-1$ state while $J\cdot n^{\rm kin}$ favors $C=+1$. Because $(J\cdot n^{\rm kin})^z$ has much larger amplitude than $h^{{\rm HF},z}$, we expect that the overall energetically favorable state has $C=+1$.

In Fig.~\ref{fig:hz_Jn}(c) we plot the $z$-component of $\bm h^{\rm HF}$ and $J\cdot n^{\rm kin}$ fields at $K$ as a function of $\gamma$. We find that the amplitude of both fields monotonically increase with $\gamma$, with the latter always several times larger than the former. In the same figure we also plot another quantity that measures the out-of-plane ferromagnetic coupling strength:
\begin{equation}
(J \cdot n^A)_K^z \equiv \sum_{\bm k} J_{\bm{Kk}}^{zz}.
\end{equation}
Physically this is the effective out-of-plane Zeeman field at $K$ produced by the $A$-sublattice polarized state $\bm{n_k}^A = (0,0,1)$. We find that the ferromagnetic coupling strength is significantly reduced as $\gamma$ increases from 0 to 1. Weakening of ferromagnetic coupling increases the energy of sublattice-polarized states and further stabilizes the AHC state that is favored by the effective pseudospin Zeeman fields.

\begin{figure*}
    \centering
    \includegraphics[width=0.27\linewidth]{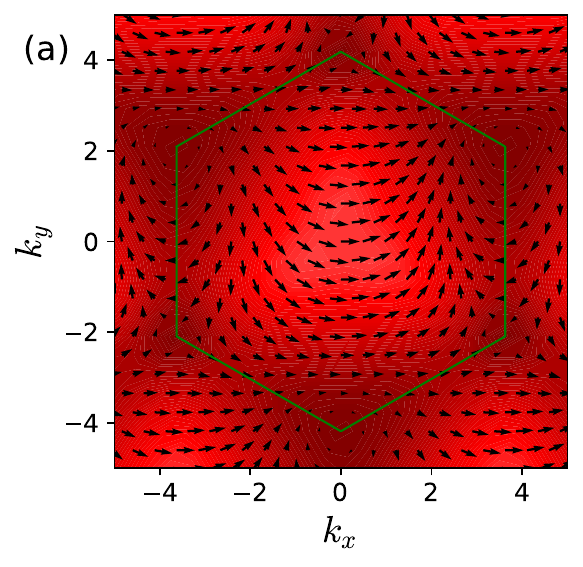}
    \includegraphics[width=0.335\linewidth]{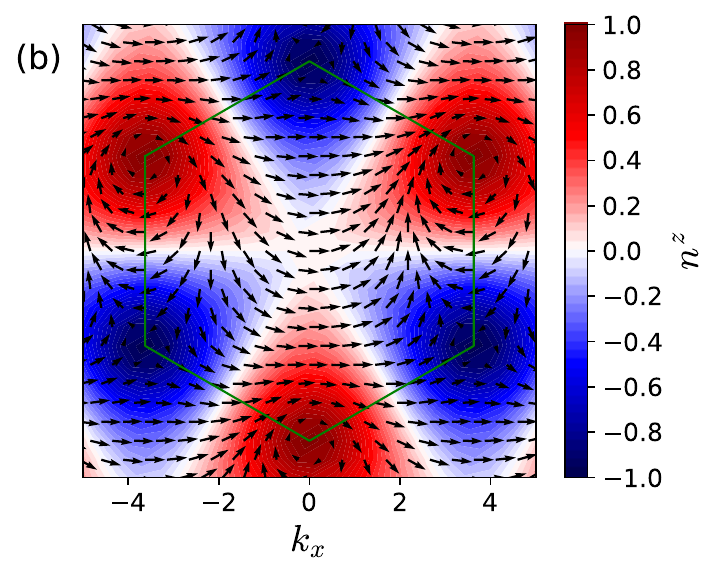}
    \includegraphics[width=0.36\linewidth]{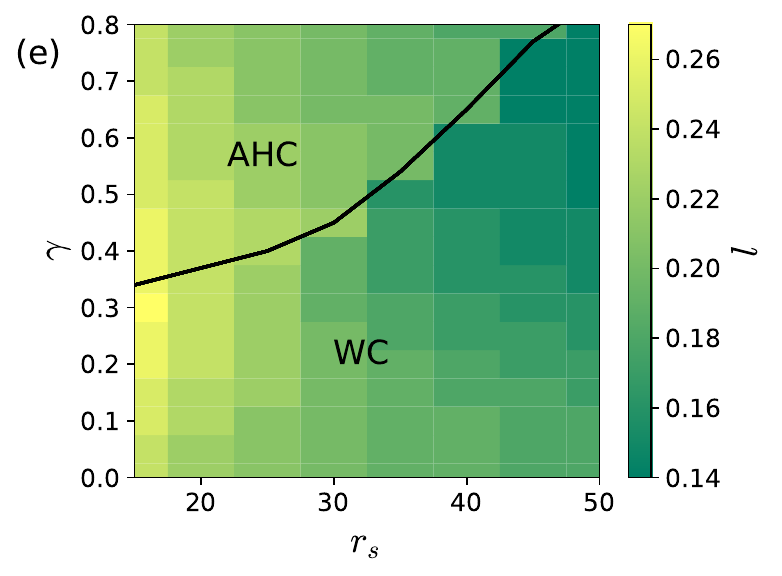}
    \includegraphics[width=0.35\linewidth]{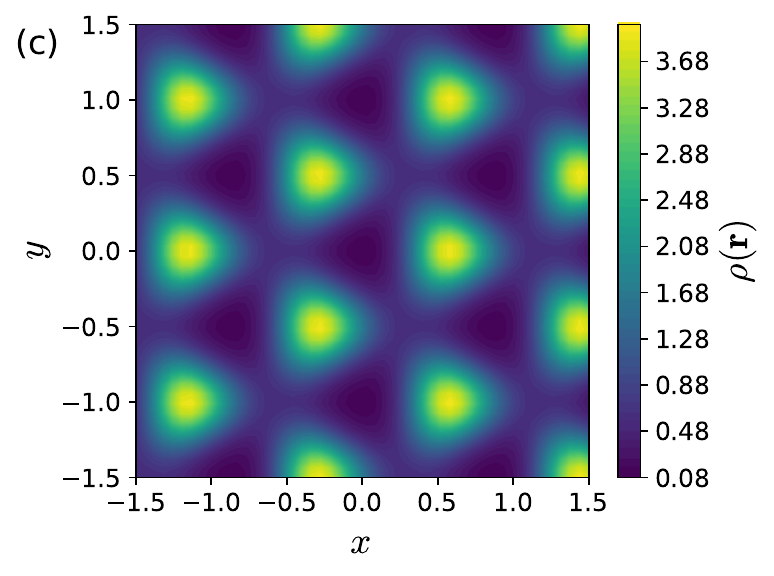}
    \includegraphics[width=0.34\linewidth]{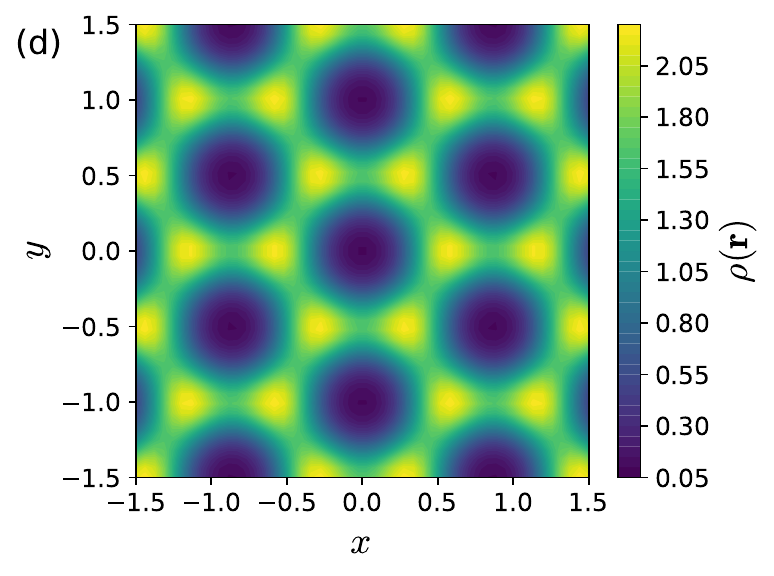}
    \includegraphics[width=0.265\linewidth]{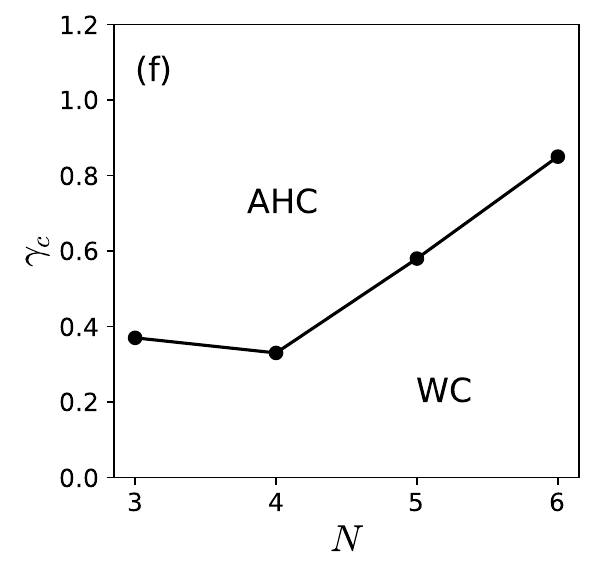}
    \caption{(a-b): Sublattice pseudospin textures of the ground states at $r_s=20$ and (a) $\gamma=0.3$; (b) $\gamma=0.4$. The winding number is $N=3$ for figures (a-e). (c-d): Charge density distribution of the states in (a-b). (e) Phase diagram as a function of $r_s$ and $\gamma$. The color scale represents the localization length $l$ of the ground state, and the black curve separates the trivial Wigner crystal (WC) and $C=1$ anomalous Hall crystal (AHC) states. (f) The critical value of $\gamma$ for different winding numbers $N$ at $r_s=20$.}
    \label{fig:phase_diagram}
\end{figure*}

{\it Phase diagram.---}
To find the mean-field ground state of the system, we project the system Hamiltonian onto the two-sublattice subspace and perform self-consistent Hartree-Fock calculations. The sublattice projection introduces an extra variational parameter $l$ that has the physical meaning of a localization length for charge localized about a particular honeycomb lattice site and that needs to be optimized. For given $r_s$ and $\gamma$, we find the self-consistent solutions at each $l$. Then we compare solutions at different $l$ and identify the one with lowest energy as the mean-field ground state.

Fig.~\ref{fig:phase_diagram}(a) and (b) show the pseudospin textures of ground states at two different values of $\gamma$, corresponding respectively to a trivial WC state and a $C=1$ AHC state. The charge density profiles in Fig.~\ref{fig:phase_diagram}(c) and (d) make it clear that the WC state forms a triangular lattice while the AHC state forms a honeycomb lattice. Fig.~\ref{fig:phase_diagram}(e) shows the phase diagram in $(r_s, \gamma)$ parameter space, where the color scale represents the localization length of the ground state. While the WC state is always the ground state when the form factor is trivial, a first-order transition to the AHC state occurs as $\gamma$ increases. The critical value of $\gamma$ increases with $r_s$.

Our sublattice-projected theory by construction does not consider melting of crystals into translationally invariant liquid phases. A useful approximate melting criterion is provided by the Lindemann criterion \cite{khrapak2020lindemann, zheng1998lindemann, bedanov1985modified, goldoni1996stability}, which in our context states that melting occurs when the localization length $l$ (in units of lattice constant $a$) reaches a critical value. The color scale in Fig.~\ref{fig:phase_diagram}(e) shows that the localization length of both WC and AHC states decreases with $r_s$. Although the precise critical value of $l$ is unknown for the AHC state, since the phase boundary moves towards large $r_s$ as $\gamma$ increases, we expect that at large $\gamma$ the AHC state is stabilized as an intermediate phase between the WC and Fermi liquid phases, leading to the schematic phase diagram in Fig.~\ref{fig:schematic_phase}.

{\it Discussion.---}
In this Letter we used a simple model to demonstrate the possibility of spontaneous crystallization of 2D electron systems into a topologically nontrivial state. Our sublattice pseudospin picture qualitatively explains the physical origin of the AHC states proposed by recent theoretical work \cite{dong2023anomalous, dong2023theory, zhou2023fractional, guo2023theory, kwan2023moire}, but goes beyond the context of multilayer graphene. Our theory shows that the most essential ingredient for AHCs is a nontrivial form factor $\Lambda_{\bm p', \bm p}$ with Berry curvature concentrated on the scale of the superlattice mBZ that breaks the effective TRS and favors a topologically nontrivial sublattice pseudospin texture. Nontrivial form factors also weaken the ferromagnetic exchange coupling that favors sublattice-polarized states and thus further stabilizes the AHC state.

An interesting open question is the optimal form of $\Lambda_{\bm p', \bm p}$ for the realization of AHCs. Our calculations at different winding numbers $N$ show that the AHC area in the phase diagram does not monotonically increase with the Berry flux in the first mBZ. Fig.~\ref{fig:phase_diagram}(f) shows the critical value of $\gamma$ for the transition from WC to AHC states at $N=3,4,5,6$ with fixed $r_s=20$. At $N=1,2$, and 7, the WC state remains the ground state up to very large $\gamma$. Analytic progress on the $h$ and $J$ coefficients (see expressions in Supplemental Material) or analogous studies using the controlled quantum geometry of ideal bands \cite{wang2021exact,estienne2023ideal,crepel2023chiral} can shed light on this non-monotonic behavior and help identify the optimal form of $\Lambda_{\bm p', \bm p}$ as well as promising material candidates for the realization of AHCs.

Since all electrons in the WC state are well localized, the WC state is well described by the Slater determinant \eqref{eq:Psi} and its total energy is largely independent of the specific form of sublattice basis orbitals we use as long as it has the correct localization length. For the AHC state, on the other hand, it is less clear whether the ansatz \eqref{eq:Psi} and the sublattice basis construction provide an accurate description; if not, our calculations overestimate the energy of AHCs. Therefore, the critical $\gamma$ in our results should be regarded as an upper bound for the realistic value. More sophisticated computational techniques such as quantum Monte Carlo methods are required to obtain a more accurate phase diagram.

{\it Acknowledgements.---}
Y.Z. thanks Allan MacDonald for very helpful discussion on pseudospin order in graphene. Y.Z. and A.J.M. acknowledge support from Programmable Quantum Materials, an Energy Frontiers Research Center funded by the U.S. Department of Energy (DOE), Office of Science, Basic Energy Sciences (BES), under award DE-SC0019443.   
J.C. acknowledges support from the Air Force Office of Scientific Research under Grant No. FA9550-20-1-0260 and is partially supported by the Alfred P. Sloan Foundation through a Sloan Research Fellowship.
The Flatiron Institute is a division of the Simons Foundation.

\bibliography{references}

\begin{thebibliography}{54}%
\makeatletter
\providecommand \@ifxundefined [1]{%
 \@ifx{#1\undefined}
}%
\providecommand \@ifnum [1]{%
 \ifnum #1\expandafter \@firstoftwo
 \else \expandafter \@secondoftwo
 \fi
}%
\providecommand \@ifx [1]{%
 \ifx #1\expandafter \@firstoftwo
 \else \expandafter \@secondoftwo
 \fi
}%
\providecommand \natexlab [1]{#1}%
\providecommand \enquote  [1]{``#1''}%
\providecommand \bibnamefont  [1]{#1}%
\providecommand \bibfnamefont [1]{#1}%
\providecommand \citenamefont [1]{#1}%
\providecommand \href@noop [0]{\@secondoftwo}%
\providecommand \href [0]{\begingroup \@sanitize@url \@href}%
\providecommand \@href[1]{\@@startlink{#1}\@@href}%
\providecommand \@@href[1]{\endgroup#1\@@endlink}%
\providecommand \@sanitize@url [0]{\catcode `\\12\catcode `\$12\catcode
  `\&12\catcode `\#12\catcode `\^12\catcode `\_12\catcode `\%12\relax}%
\providecommand \@@startlink[1]{}%
\providecommand \@@endlink[0]{}%
\providecommand \url  [0]{\begingroup\@sanitize@url \@url }%
\providecommand \@url [1]{\endgroup\@href {#1}{\urlprefix }}%
\providecommand \urlprefix  [0]{URL }%
\providecommand \Eprint [0]{\href }%
\providecommand \doibase [0]{https://doi.org/}%
\providecommand \selectlanguage [0]{\@gobble}%
\providecommand \bibinfo  [0]{\@secondoftwo}%
\providecommand \bibfield  [0]{\@secondoftwo}%
\providecommand \translation [1]{[#1]}%
\providecommand \BibitemOpen [0]{}%
\providecommand \bibitemStop [0]{}%
\providecommand \bibitemNoStop [0]{.\EOS\space}%
\providecommand \EOS [0]{\spacefactor3000\relax}%
\providecommand \BibitemShut  [1]{\csname bibitem#1\endcsname}%
\let\auto@bib@innerbib\@empty
\bibitem [{\citenamefont {Wigner}(1934)}]{wigner1934}%
  \BibitemOpen
  \bibfield  {author} {\bibinfo {author} {\bibfnamefont {E.}~\bibnamefont
  {Wigner}},\ }\bibfield  {title} {\bibinfo {title} {On the interaction of
  electrons in metals},\ }\href {https://doi.org/10.1103/PhysRev.46.1002}
  {\bibfield  {journal} {\bibinfo  {journal} {Phys. Rev.}\ }\textbf {\bibinfo
  {volume} {46}},\ \bibinfo {pages} {1002} (\bibinfo {year}
  {1934})}\BibitemShut {NoStop}%
\bibitem [{\citenamefont {Bonsall}\ and\ \citenamefont
  {Maradudin}(1977)}]{bonsall1977static}%
  \BibitemOpen
  \bibfield  {author} {\bibinfo {author} {\bibfnamefont {L.}~\bibnamefont
  {Bonsall}}\ and\ \bibinfo {author} {\bibfnamefont {A.~A.}\ \bibnamefont
  {Maradudin}},\ }\bibfield  {title} {\bibinfo {title} {Some static and
  dynamical properties of a two-dimensional {Wigner} crystal},\ }\href
  {https://doi.org/10.1103/PhysRevB.15.1959} {\bibfield  {journal} {\bibinfo
  {journal} {Phys. Rev. B}\ }\textbf {\bibinfo {volume} {15}},\ \bibinfo
  {pages} {1959} (\bibinfo {year} {1977})}\BibitemShut {NoStop}%
\bibitem [{\citenamefont {Tanatar}\ and\ \citenamefont
  {Ceperley}(1989)}]{tanatar1989ground}%
  \BibitemOpen
  \bibfield  {author} {\bibinfo {author} {\bibfnamefont {B.}~\bibnamefont
  {Tanatar}}\ and\ \bibinfo {author} {\bibfnamefont {D.~M.}\ \bibnamefont
  {Ceperley}},\ }\bibfield  {title} {\bibinfo {title} {Ground state of the
  two-dimensional electron gas},\ }\href
  {https://doi.org/10.1103/PhysRevB.39.5005} {\bibfield  {journal} {\bibinfo
  {journal} {Phys. Rev. B}\ }\textbf {\bibinfo {volume} {39}},\ \bibinfo
  {pages} {5005} (\bibinfo {year} {1989})}\BibitemShut {NoStop}%
\bibitem [{\citenamefont {Drummond}\ and\ \citenamefont
  {Needs}(2009)}]{drummond2009phase}%
  \BibitemOpen
  \bibfield  {author} {\bibinfo {author} {\bibfnamefont {N.~D.}\ \bibnamefont
  {Drummond}}\ and\ \bibinfo {author} {\bibfnamefont {R.~J.}\ \bibnamefont
  {Needs}},\ }\bibfield  {title} {\bibinfo {title} {Phase diagram of the
  low-density two-dimensional homogeneous electron gas},\ }\href
  {https://doi.org/10.1103/PhysRevLett.102.126402} {\bibfield  {journal}
  {\bibinfo  {journal} {Phys. Rev. Lett.}\ }\textbf {\bibinfo {volume} {102}},\
  \bibinfo {pages} {126402} (\bibinfo {year} {2009})}\BibitemShut {NoStop}%
\bibitem [{\citenamefont {Spivak}\ and\ \citenamefont
  {Kivelson}(2004)}]{spivak2004phases}%
  \BibitemOpen
  \bibfield  {author} {\bibinfo {author} {\bibfnamefont {B.}~\bibnamefont
  {Spivak}}\ and\ \bibinfo {author} {\bibfnamefont {S.~A.}\ \bibnamefont
  {Kivelson}},\ }\bibfield  {title} {\bibinfo {title} {Phases intermediate
  between a two-dimensional electron liquid and wigner crystal},\ }\href
  {https://doi.org/10.1103/PhysRevB.70.155114} {\bibfield  {journal} {\bibinfo
  {journal} {Phys. Rev. B}\ }\textbf {\bibinfo {volume} {70}},\ \bibinfo
  {pages} {155114} (\bibinfo {year} {2004})}\BibitemShut {NoStop}%
\bibitem [{\citenamefont {Zarenia}\ \emph {et~al.}(2017)\citenamefont
  {Zarenia}, \citenamefont {Neilson}, \citenamefont {Partoens},\ and\
  \citenamefont {Peeters}}]{zarenia2017wigner}%
  \BibitemOpen
  \bibfield  {author} {\bibinfo {author} {\bibfnamefont {M.}~\bibnamefont
  {Zarenia}}, \bibinfo {author} {\bibfnamefont {D.}~\bibnamefont {Neilson}},
  \bibinfo {author} {\bibfnamefont {B.}~\bibnamefont {Partoens}},\ and\
  \bibinfo {author} {\bibfnamefont {F.~M.}\ \bibnamefont {Peeters}},\
  }\bibfield  {title} {\bibinfo {title} {Wigner crystallization in transition
  metal dichalcogenides: A new approach to correlation energy},\ }\href
  {https://doi.org/10.1103/PhysRevB.95.115438} {\bibfield  {journal} {\bibinfo
  {journal} {Phys. Rev. B}\ }\textbf {\bibinfo {volume} {95}},\ \bibinfo
  {pages} {115438} (\bibinfo {year} {2017})}\BibitemShut {NoStop}%
\bibitem [{\citenamefont {Goldman}\ \emph {et~al.}(1990)\citenamefont
  {Goldman}, \citenamefont {Santos}, \citenamefont {Shayegan},\ and\
  \citenamefont {Cunningham}}]{goldman1990evidence}%
  \BibitemOpen
  \bibfield  {author} {\bibinfo {author} {\bibfnamefont {V.~J.}\ \bibnamefont
  {Goldman}}, \bibinfo {author} {\bibfnamefont {M.}~\bibnamefont {Santos}},
  \bibinfo {author} {\bibfnamefont {M.}~\bibnamefont {Shayegan}},\ and\
  \bibinfo {author} {\bibfnamefont {J.~E.}\ \bibnamefont {Cunningham}},\
  }\bibfield  {title} {\bibinfo {title} {Evidence for two-dimentional quantum
  wigner crystal},\ }\href {https://doi.org/10.1103/PhysRevLett.65.2189}
  {\bibfield  {journal} {\bibinfo  {journal} {Phys. Rev. Lett.}\ }\textbf
  {\bibinfo {volume} {65}},\ \bibinfo {pages} {2189} (\bibinfo {year}
  {1990})}\BibitemShut {NoStop}%
\bibitem [{\citenamefont {Yoon}\ \emph {et~al.}(1999)\citenamefont {Yoon},
  \citenamefont {Li}, \citenamefont {Shahar}, \citenamefont {Tsui},\ and\
  \citenamefont {Shayegan}}]{yoon1999wigner}%
  \BibitemOpen
  \bibfield  {author} {\bibinfo {author} {\bibfnamefont {J.}~\bibnamefont
  {Yoon}}, \bibinfo {author} {\bibfnamefont {C.~C.}\ \bibnamefont {Li}},
  \bibinfo {author} {\bibfnamefont {D.}~\bibnamefont {Shahar}}, \bibinfo
  {author} {\bibfnamefont {D.~C.}\ \bibnamefont {Tsui}},\ and\ \bibinfo
  {author} {\bibfnamefont {M.}~\bibnamefont {Shayegan}},\ }\bibfield  {title}
  {\bibinfo {title} {Wigner crystallization and metal-insulator transition of
  two-dimensional holes in gaas at
  $\mathit{B}\phantom{\rule{0ex}{0ex}}=\phantom{\rule{0ex}{0ex}}0$},\ }\href
  {https://doi.org/10.1103/PhysRevLett.82.1744} {\bibfield  {journal} {\bibinfo
   {journal} {Phys. Rev. Lett.}\ }\textbf {\bibinfo {volume} {82}},\ \bibinfo
  {pages} {1744} (\bibinfo {year} {1999})}\BibitemShut {NoStop}%
\bibitem [{\citenamefont {Hossain}\ \emph {et~al.}(2020)\citenamefont
  {Hossain}, \citenamefont {Ma}, \citenamefont {Rosales}, \citenamefont
  {Chung}, \citenamefont {Pfeiffer}, \citenamefont {West}, \citenamefont
  {Baldwin},\ and\ \citenamefont {Shayegan}}]{hossain2020observation}%
  \BibitemOpen
  \bibfield  {author} {\bibinfo {author} {\bibfnamefont {M.~S.}\ \bibnamefont
  {Hossain}}, \bibinfo {author} {\bibfnamefont {M.}~\bibnamefont {Ma}},
  \bibinfo {author} {\bibfnamefont {K.~V.}\ \bibnamefont {Rosales}}, \bibinfo
  {author} {\bibfnamefont {Y.}~\bibnamefont {Chung}}, \bibinfo {author}
  {\bibfnamefont {L.}~\bibnamefont {Pfeiffer}}, \bibinfo {author}
  {\bibfnamefont {K.}~\bibnamefont {West}}, \bibinfo {author} {\bibfnamefont
  {K.}~\bibnamefont {Baldwin}},\ and\ \bibinfo {author} {\bibfnamefont
  {M.}~\bibnamefont {Shayegan}},\ }\bibfield  {title} {\bibinfo {title}
  {Observation of spontaneous ferromagnetism in a two-dimensional electron
  system},\ }\href@noop {} {\bibfield  {journal} {\bibinfo  {journal}
  {Proceedings of the National Academy of Sciences}\ }\textbf {\bibinfo
  {volume} {117}},\ \bibinfo {pages} {32244} (\bibinfo {year}
  {2020})}\BibitemShut {NoStop}%
\bibitem [{\citenamefont {Zhou}\ \emph {et~al.}(2021)\citenamefont {Zhou},
  \citenamefont {Sung}, \citenamefont {Brutschea}, \citenamefont {Esterlis},
  \citenamefont {Wang}, \citenamefont {Scuri}, \citenamefont {Gelly},
  \citenamefont {Heo}, \citenamefont {Taniguchi}, \citenamefont {Watanabe}
  \emph {et~al.}}]{zhou2021bilayer}%
  \BibitemOpen
  \bibfield  {author} {\bibinfo {author} {\bibfnamefont {Y.}~\bibnamefont
  {Zhou}}, \bibinfo {author} {\bibfnamefont {J.}~\bibnamefont {Sung}}, \bibinfo
  {author} {\bibfnamefont {E.}~\bibnamefont {Brutschea}}, \bibinfo {author}
  {\bibfnamefont {I.}~\bibnamefont {Esterlis}}, \bibinfo {author}
  {\bibfnamefont {Y.}~\bibnamefont {Wang}}, \bibinfo {author} {\bibfnamefont
  {G.}~\bibnamefont {Scuri}}, \bibinfo {author} {\bibfnamefont {R.~J.}\
  \bibnamefont {Gelly}}, \bibinfo {author} {\bibfnamefont {H.}~\bibnamefont
  {Heo}}, \bibinfo {author} {\bibfnamefont {T.}~\bibnamefont {Taniguchi}},
  \bibinfo {author} {\bibfnamefont {K.}~\bibnamefont {Watanabe}}, \emph
  {et~al.},\ }\bibfield  {title} {\bibinfo {title} {Bilayer wigner crystals in
  a transition metal dichalcogenide heterostructure},\ }\href@noop {}
  {\bibfield  {journal} {\bibinfo  {journal} {Nature}\ }\textbf {\bibinfo
  {volume} {595}},\ \bibinfo {pages} {48} (\bibinfo {year} {2021})}\BibitemShut
  {NoStop}%
\bibitem [{\citenamefont {Smole{\'n}ski}\ \emph {et~al.}(2021)\citenamefont
  {Smole{\'n}ski}, \citenamefont {Dolgirev}, \citenamefont {Kuhlenkamp},
  \citenamefont {Popert}, \citenamefont {Shimazaki}, \citenamefont {Back},
  \citenamefont {Lu}, \citenamefont {Kroner}, \citenamefont {Watanabe},
  \citenamefont {Taniguchi} \emph {et~al.}}]{smolenski2021signatures}%
  \BibitemOpen
  \bibfield  {author} {\bibinfo {author} {\bibfnamefont {T.}~\bibnamefont
  {Smole{\'n}ski}}, \bibinfo {author} {\bibfnamefont {P.~E.}\ \bibnamefont
  {Dolgirev}}, \bibinfo {author} {\bibfnamefont {C.}~\bibnamefont
  {Kuhlenkamp}}, \bibinfo {author} {\bibfnamefont {A.}~\bibnamefont {Popert}},
  \bibinfo {author} {\bibfnamefont {Y.}~\bibnamefont {Shimazaki}}, \bibinfo
  {author} {\bibfnamefont {P.}~\bibnamefont {Back}}, \bibinfo {author}
  {\bibfnamefont {X.}~\bibnamefont {Lu}}, \bibinfo {author} {\bibfnamefont
  {M.}~\bibnamefont {Kroner}}, \bibinfo {author} {\bibfnamefont
  {K.}~\bibnamefont {Watanabe}}, \bibinfo {author} {\bibfnamefont
  {T.}~\bibnamefont {Taniguchi}}, \emph {et~al.},\ }\bibfield  {title}
  {\bibinfo {title} {Signatures of wigner crystal of electrons in a monolayer
  semiconductor},\ }\href@noop {} {\bibfield  {journal} {\bibinfo  {journal}
  {Nature}\ }\textbf {\bibinfo {volume} {595}},\ \bibinfo {pages} {53}
  (\bibinfo {year} {2021})}\BibitemShut {NoStop}%
\bibitem [{\citenamefont {Li}\ \emph {et~al.}(2021{\natexlab{a}})\citenamefont
  {Li}, \citenamefont {Li}, \citenamefont {Regan}, \citenamefont {Wang},
  \citenamefont {Zhao}, \citenamefont {Kahn}, \citenamefont {Yumigeta},
  \citenamefont {Blei}, \citenamefont {Taniguchi}, \citenamefont {Watanabe}
  \emph {et~al.}}]{li2021imaging}%
  \BibitemOpen
  \bibfield  {author} {\bibinfo {author} {\bibfnamefont {H.}~\bibnamefont
  {Li}}, \bibinfo {author} {\bibfnamefont {S.}~\bibnamefont {Li}}, \bibinfo
  {author} {\bibfnamefont {E.~C.}\ \bibnamefont {Regan}}, \bibinfo {author}
  {\bibfnamefont {D.}~\bibnamefont {Wang}}, \bibinfo {author} {\bibfnamefont
  {W.}~\bibnamefont {Zhao}}, \bibinfo {author} {\bibfnamefont {S.}~\bibnamefont
  {Kahn}}, \bibinfo {author} {\bibfnamefont {K.}~\bibnamefont {Yumigeta}},
  \bibinfo {author} {\bibfnamefont {M.}~\bibnamefont {Blei}}, \bibinfo {author}
  {\bibfnamefont {T.}~\bibnamefont {Taniguchi}}, \bibinfo {author}
  {\bibfnamefont {K.}~\bibnamefont {Watanabe}}, \emph {et~al.},\ }\bibfield
  {title} {\bibinfo {title} {Imaging two-dimensional generalized wigner
  crystals},\ }\href@noop {} {\bibfield  {journal} {\bibinfo  {journal}
  {Nature}\ }\textbf {\bibinfo {volume} {597}},\ \bibinfo {pages} {650}
  (\bibinfo {year} {2021}{\natexlab{a}})}\BibitemShut {NoStop}%
\bibitem [{\citenamefont {Regan}\ \emph {et~al.}(2020)\citenamefont {Regan},
  \citenamefont {Wang}, \citenamefont {Jin}, \citenamefont {Bakti~Utama},
  \citenamefont {Gao}, \citenamefont {Wei}, \citenamefont {Zhao}, \citenamefont
  {Zhao}, \citenamefont {Zhang}, \citenamefont {Yumigeta} \emph
  {et~al.}}]{regan2020mott}%
  \BibitemOpen
  \bibfield  {author} {\bibinfo {author} {\bibfnamefont {E.~C.}\ \bibnamefont
  {Regan}}, \bibinfo {author} {\bibfnamefont {D.}~\bibnamefont {Wang}},
  \bibinfo {author} {\bibfnamefont {C.}~\bibnamefont {Jin}}, \bibinfo {author}
  {\bibfnamefont {M.~I.}\ \bibnamefont {Bakti~Utama}}, \bibinfo {author}
  {\bibfnamefont {B.}~\bibnamefont {Gao}}, \bibinfo {author} {\bibfnamefont
  {X.}~\bibnamefont {Wei}}, \bibinfo {author} {\bibfnamefont {S.}~\bibnamefont
  {Zhao}}, \bibinfo {author} {\bibfnamefont {W.}~\bibnamefont {Zhao}}, \bibinfo
  {author} {\bibfnamefont {Z.}~\bibnamefont {Zhang}}, \bibinfo {author}
  {\bibfnamefont {K.}~\bibnamefont {Yumigeta}}, \emph {et~al.},\ }\bibfield
  {title} {\bibinfo {title} {{Mott and generalized Wigner crystal states in
  WSe$_2$/WS$_2$ moir{\'e} superlattices}},\ }\href@noop {} {\bibfield
  {journal} {\bibinfo  {journal} {Nature}\ }\textbf {\bibinfo {volume} {579}},\
  \bibinfo {pages} {359} (\bibinfo {year} {2020})}\BibitemShut {NoStop}%
\bibitem [{\citenamefont {Xiang}\ \emph {et~al.}(2024)\citenamefont {Xiang},
  \citenamefont {Li}, \citenamefont {Xiao}, \citenamefont {Naik}, \citenamefont
  {Ge}, \citenamefont {He}, \citenamefont {Chen}, \citenamefont {Nie},
  \citenamefont {Li}, \citenamefont {Jiang} \emph {et~al.}}]{xiang2024quantum}%
  \BibitemOpen
  \bibfield  {author} {\bibinfo {author} {\bibfnamefont {Z.}~\bibnamefont
  {Xiang}}, \bibinfo {author} {\bibfnamefont {H.}~\bibnamefont {Li}}, \bibinfo
  {author} {\bibfnamefont {J.}~\bibnamefont {Xiao}}, \bibinfo {author}
  {\bibfnamefont {M.~H.}\ \bibnamefont {Naik}}, \bibinfo {author}
  {\bibfnamefont {Z.}~\bibnamefont {Ge}}, \bibinfo {author} {\bibfnamefont
  {Z.}~\bibnamefont {He}}, \bibinfo {author} {\bibfnamefont {S.}~\bibnamefont
  {Chen}}, \bibinfo {author} {\bibfnamefont {J.}~\bibnamefont {Nie}}, \bibinfo
  {author} {\bibfnamefont {S.}~\bibnamefont {Li}}, \bibinfo {author}
  {\bibfnamefont {Y.}~\bibnamefont {Jiang}}, \emph {et~al.},\ }\bibfield
  {title} {\bibinfo {title} {Quantum melting of a disordered wigner solid},\
  }\href@noop {} {\bibfield  {journal} {\bibinfo  {journal} {arXiv preprint
  arXiv:2402.05456}\ } (\bibinfo {year} {2024})}\BibitemShut {NoStop}%
\bibitem [{\citenamefont {Chang}\ \emph {et~al.}(2023)\citenamefont {Chang},
  \citenamefont {Liu},\ and\ \citenamefont {MacDonald}}]{chang2023colloquium}%
  \BibitemOpen
  \bibfield  {author} {\bibinfo {author} {\bibfnamefont {C.-Z.}\ \bibnamefont
  {Chang}}, \bibinfo {author} {\bibfnamefont {C.-X.}\ \bibnamefont {Liu}},\
  and\ \bibinfo {author} {\bibfnamefont {A.~H.}\ \bibnamefont {MacDonald}},\
  }\bibfield  {title} {\bibinfo {title} {Colloquium: Quantum anomalous hall
  effect},\ }\href {https://doi.org/10.1103/RevModPhys.95.011002} {\bibfield
  {journal} {\bibinfo  {journal} {Rev. Mod. Phys.}\ }\textbf {\bibinfo {volume}
  {95}},\ \bibinfo {pages} {011002} (\bibinfo {year} {2023})}\BibitemShut
  {NoStop}%
\bibitem [{\citenamefont {Haldane}(1988)}]{haldane1988model}%
  \BibitemOpen
  \bibfield  {author} {\bibinfo {author} {\bibfnamefont {F.~D.~M.}\
  \bibnamefont {Haldane}},\ }\bibfield  {title} {\bibinfo {title} {Model for a
  quantum hall effect without landau levels: Condensed-matter realization of
  the ``parity anomaly"},\ }\href {https://doi.org/10.1103/PhysRevLett.61.2015}
  {\bibfield  {journal} {\bibinfo  {journal} {Phys. Rev. Lett.}\ }\textbf
  {\bibinfo {volume} {61}},\ \bibinfo {pages} {2015} (\bibinfo {year}
  {1988})}\BibitemShut {NoStop}%
\bibitem [{\citenamefont {Chang}\ \emph {et~al.}(2013)\citenamefont {Chang},
  \citenamefont {Zhang}, \citenamefont {Feng}, \citenamefont {Shen},
  \citenamefont {Zhang}, \citenamefont {Guo}, \citenamefont {Li}, \citenamefont
  {Ou}, \citenamefont {Wei}, \citenamefont {Wang} \emph
  {et~al.}}]{chang2013experimental}%
  \BibitemOpen
  \bibfield  {author} {\bibinfo {author} {\bibfnamefont {C.-Z.}\ \bibnamefont
  {Chang}}, \bibinfo {author} {\bibfnamefont {J.}~\bibnamefont {Zhang}},
  \bibinfo {author} {\bibfnamefont {X.}~\bibnamefont {Feng}}, \bibinfo {author}
  {\bibfnamefont {J.}~\bibnamefont {Shen}}, \bibinfo {author} {\bibfnamefont
  {Z.}~\bibnamefont {Zhang}}, \bibinfo {author} {\bibfnamefont
  {M.}~\bibnamefont {Guo}}, \bibinfo {author} {\bibfnamefont {K.}~\bibnamefont
  {Li}}, \bibinfo {author} {\bibfnamefont {Y.}~\bibnamefont {Ou}}, \bibinfo
  {author} {\bibfnamefont {P.}~\bibnamefont {Wei}}, \bibinfo {author}
  {\bibfnamefont {L.-L.}\ \bibnamefont {Wang}}, \emph {et~al.},\ }\bibfield
  {title} {\bibinfo {title} {Experimental observation of the quantum anomalous
  hall effect in a magnetic topological insulator},\ }\href@noop {} {\bibfield
  {journal} {\bibinfo  {journal} {Science}\ }\textbf {\bibinfo {volume}
  {340}},\ \bibinfo {pages} {167} (\bibinfo {year} {2013})}\BibitemShut
  {NoStop}%
\bibitem [{\citenamefont {Chang}\ \emph {et~al.}(2015)\citenamefont {Chang},
  \citenamefont {Zhao}, \citenamefont {Kim}, \citenamefont {Zhang},
  \citenamefont {Assaf}, \citenamefont {Heiman}, \citenamefont {Zhang},
  \citenamefont {Liu}, \citenamefont {Chan},\ and\ \citenamefont
  {Moodera}}]{chang2015high}%
  \BibitemOpen
  \bibfield  {author} {\bibinfo {author} {\bibfnamefont {C.-Z.}\ \bibnamefont
  {Chang}}, \bibinfo {author} {\bibfnamefont {W.}~\bibnamefont {Zhao}},
  \bibinfo {author} {\bibfnamefont {D.~Y.}\ \bibnamefont {Kim}}, \bibinfo
  {author} {\bibfnamefont {H.}~\bibnamefont {Zhang}}, \bibinfo {author}
  {\bibfnamefont {B.~A.}\ \bibnamefont {Assaf}}, \bibinfo {author}
  {\bibfnamefont {D.}~\bibnamefont {Heiman}}, \bibinfo {author} {\bibfnamefont
  {S.-C.}\ \bibnamefont {Zhang}}, \bibinfo {author} {\bibfnamefont
  {C.}~\bibnamefont {Liu}}, \bibinfo {author} {\bibfnamefont {M.~H.}\
  \bibnamefont {Chan}},\ and\ \bibinfo {author} {\bibfnamefont {J.~S.}\
  \bibnamefont {Moodera}},\ }\bibfield  {title} {\bibinfo {title}
  {High-precision realization of robust quantum anomalous hall state in a hard
  ferromagnetic topological insulator},\ }\href@noop {} {\bibfield  {journal}
  {\bibinfo  {journal} {Nature materials}\ }\textbf {\bibinfo {volume} {14}},\
  \bibinfo {pages} {473} (\bibinfo {year} {2015})}\BibitemShut {NoStop}%
\bibitem [{\citenamefont {Deng}\ \emph {et~al.}(2020)\citenamefont {Deng},
  \citenamefont {Yu}, \citenamefont {Shi}, \citenamefont {Guo}, \citenamefont
  {Xu}, \citenamefont {Wang}, \citenamefont {Chen},\ and\ \citenamefont
  {Zhang}}]{deng2020quantum}%
  \BibitemOpen
  \bibfield  {author} {\bibinfo {author} {\bibfnamefont {Y.}~\bibnamefont
  {Deng}}, \bibinfo {author} {\bibfnamefont {Y.}~\bibnamefont {Yu}}, \bibinfo
  {author} {\bibfnamefont {M.~Z.}\ \bibnamefont {Shi}}, \bibinfo {author}
  {\bibfnamefont {Z.}~\bibnamefont {Guo}}, \bibinfo {author} {\bibfnamefont
  {Z.}~\bibnamefont {Xu}}, \bibinfo {author} {\bibfnamefont {J.}~\bibnamefont
  {Wang}}, \bibinfo {author} {\bibfnamefont {X.~H.}\ \bibnamefont {Chen}},\
  and\ \bibinfo {author} {\bibfnamefont {Y.}~\bibnamefont {Zhang}},\ }\bibfield
   {title} {\bibinfo {title} {{Quantum anomalous Hall effect in intrinsic
  magnetic topological insulator MnBi$_2$Te$_4$}},\ }\href@noop {} {\bibfield
  {journal} {\bibinfo  {journal} {Science}\ }\textbf {\bibinfo {volume}
  {367}},\ \bibinfo {pages} {895} (\bibinfo {year} {2020})}\BibitemShut
  {NoStop}%
\bibitem [{\citenamefont {Li}\ \emph {et~al.}(2021{\natexlab{b}})\citenamefont
  {Li}, \citenamefont {Jiang}, \citenamefont {Shen}, \citenamefont {Zhang},
  \citenamefont {Li}, \citenamefont {Tao}, \citenamefont {Devakul},
  \citenamefont {Watanabe}, \citenamefont {Taniguchi}, \citenamefont {Fu} \emph
  {et~al.}}]{li2021quantum}%
  \BibitemOpen
  \bibfield  {author} {\bibinfo {author} {\bibfnamefont {T.}~\bibnamefont
  {Li}}, \bibinfo {author} {\bibfnamefont {S.}~\bibnamefont {Jiang}}, \bibinfo
  {author} {\bibfnamefont {B.}~\bibnamefont {Shen}}, \bibinfo {author}
  {\bibfnamefont {Y.}~\bibnamefont {Zhang}}, \bibinfo {author} {\bibfnamefont
  {L.}~\bibnamefont {Li}}, \bibinfo {author} {\bibfnamefont {Z.}~\bibnamefont
  {Tao}}, \bibinfo {author} {\bibfnamefont {T.}~\bibnamefont {Devakul}},
  \bibinfo {author} {\bibfnamefont {K.}~\bibnamefont {Watanabe}}, \bibinfo
  {author} {\bibfnamefont {T.}~\bibnamefont {Taniguchi}}, \bibinfo {author}
  {\bibfnamefont {L.}~\bibnamefont {Fu}}, \emph {et~al.},\ }\bibfield  {title}
  {\bibinfo {title} {Quantum anomalous hall effect from intertwined moir{\'e}
  bands},\ }\href@noop {} {\bibfield  {journal} {\bibinfo  {journal} {Nature}\
  }\textbf {\bibinfo {volume} {600}},\ \bibinfo {pages} {641} (\bibinfo {year}
  {2021}{\natexlab{b}})}\BibitemShut {NoStop}%
\bibitem [{\citenamefont {Tao}\ \emph {et~al.}(2024)\citenamefont {Tao},
  \citenamefont {Shen}, \citenamefont {Jiang}, \citenamefont {Li},
  \citenamefont {Li}, \citenamefont {Ma}, \citenamefont {Zhao}, \citenamefont
  {Hu}, \citenamefont {Pistunova}, \citenamefont {Watanabe}, \citenamefont
  {Taniguchi}, \citenamefont {Heinz}, \citenamefont {Mak},\ and\ \citenamefont
  {Shan}}]{tao2024valley}%
  \BibitemOpen
  \bibfield  {author} {\bibinfo {author} {\bibfnamefont {Z.}~\bibnamefont
  {Tao}}, \bibinfo {author} {\bibfnamefont {B.}~\bibnamefont {Shen}}, \bibinfo
  {author} {\bibfnamefont {S.}~\bibnamefont {Jiang}}, \bibinfo {author}
  {\bibfnamefont {T.}~\bibnamefont {Li}}, \bibinfo {author} {\bibfnamefont
  {L.}~\bibnamefont {Li}}, \bibinfo {author} {\bibfnamefont {L.}~\bibnamefont
  {Ma}}, \bibinfo {author} {\bibfnamefont {W.}~\bibnamefont {Zhao}}, \bibinfo
  {author} {\bibfnamefont {J.}~\bibnamefont {Hu}}, \bibinfo {author}
  {\bibfnamefont {K.}~\bibnamefont {Pistunova}}, \bibinfo {author}
  {\bibfnamefont {K.}~\bibnamefont {Watanabe}}, \bibinfo {author}
  {\bibfnamefont {T.}~\bibnamefont {Taniguchi}}, \bibinfo {author}
  {\bibfnamefont {T.~F.}\ \bibnamefont {Heinz}}, \bibinfo {author}
  {\bibfnamefont {K.~F.}\ \bibnamefont {Mak}},\ and\ \bibinfo {author}
  {\bibfnamefont {J.}~\bibnamefont {Shan}},\ }\bibfield  {title} {\bibinfo
  {title} {{Valley-Coherent Quantum Anomalous Hall State in AB-Stacked
  ${\mathrm{MoTe}}_{2}/{\mathrm{W}\mathrm{S}\mathrm{e}}_{2}$ Bilayers}},\
  }\href {https://doi.org/10.1103/PhysRevX.14.011004} {\bibfield  {journal}
  {\bibinfo  {journal} {Phys. Rev. X}\ }\textbf {\bibinfo {volume} {14}},\
  \bibinfo {pages} {011004} (\bibinfo {year} {2024})}\BibitemShut {NoStop}%
\bibitem [{\citenamefont {Serlin}\ \emph {et~al.}(2020)\citenamefont {Serlin},
  \citenamefont {Tschirhart}, \citenamefont {Polshyn}, \citenamefont {Zhang},
  \citenamefont {Zhu}, \citenamefont {Watanabe}, \citenamefont {Taniguchi},
  \citenamefont {Balents},\ and\ \citenamefont {Young}}]{serlin2020intrinsic}%
  \BibitemOpen
  \bibfield  {author} {\bibinfo {author} {\bibfnamefont {M.}~\bibnamefont
  {Serlin}}, \bibinfo {author} {\bibfnamefont {C.}~\bibnamefont {Tschirhart}},
  \bibinfo {author} {\bibfnamefont {H.}~\bibnamefont {Polshyn}}, \bibinfo
  {author} {\bibfnamefont {Y.}~\bibnamefont {Zhang}}, \bibinfo {author}
  {\bibfnamefont {J.}~\bibnamefont {Zhu}}, \bibinfo {author} {\bibfnamefont
  {K.}~\bibnamefont {Watanabe}}, \bibinfo {author} {\bibfnamefont
  {T.}~\bibnamefont {Taniguchi}}, \bibinfo {author} {\bibfnamefont
  {L.}~\bibnamefont {Balents}},\ and\ \bibinfo {author} {\bibfnamefont
  {A.}~\bibnamefont {Young}},\ }\bibfield  {title} {\bibinfo {title} {Intrinsic
  quantized anomalous hall effect in a moir{\'e} heterostructure},\ }\href@noop
  {} {\bibfield  {journal} {\bibinfo  {journal} {Science}\ }\textbf {\bibinfo
  {volume} {367}},\ \bibinfo {pages} {900} (\bibinfo {year}
  {2020})}\BibitemShut {NoStop}%
\bibitem [{\citenamefont {Han}\ \emph {et~al.}(2023{\natexlab{a}})\citenamefont
  {Han}, \citenamefont {Lu}, \citenamefont {Scuri}, \citenamefont {Sung},
  \citenamefont {Wang}, \citenamefont {Han}, \citenamefont {Watanabe},
  \citenamefont {Taniguchi}, \citenamefont {Park},\ and\ \citenamefont
  {Ju}}]{han2023correlated}%
  \BibitemOpen
  \bibfield  {author} {\bibinfo {author} {\bibfnamefont {T.}~\bibnamefont
  {Han}}, \bibinfo {author} {\bibfnamefont {Z.}~\bibnamefont {Lu}}, \bibinfo
  {author} {\bibfnamefont {G.}~\bibnamefont {Scuri}}, \bibinfo {author}
  {\bibfnamefont {J.}~\bibnamefont {Sung}}, \bibinfo {author} {\bibfnamefont
  {J.}~\bibnamefont {Wang}}, \bibinfo {author} {\bibfnamefont {T.}~\bibnamefont
  {Han}}, \bibinfo {author} {\bibfnamefont {K.}~\bibnamefont {Watanabe}},
  \bibinfo {author} {\bibfnamefont {T.}~\bibnamefont {Taniguchi}}, \bibinfo
  {author} {\bibfnamefont {H.}~\bibnamefont {Park}},\ and\ \bibinfo {author}
  {\bibfnamefont {L.}~\bibnamefont {Ju}},\ }\bibfield  {title} {\bibinfo
  {title} {Correlated insulator and chern insulators in pentalayer rhombohedral
  stacked graphene},\ }\href@noop {} {\bibfield  {journal} {\bibinfo  {journal}
  {arXiv preprint arXiv:2305.03151}\ } (\bibinfo {year}
  {2023}{\natexlab{a}})}\BibitemShut {NoStop}%
\bibitem [{\citenamefont {Lu}\ \emph {et~al.}(2023)\citenamefont {Lu},
  \citenamefont {Han}, \citenamefont {Yao}, \citenamefont {Reddy},
  \citenamefont {Yang}, \citenamefont {Seo}, \citenamefont {Watanabe},
  \citenamefont {Taniguchi}, \citenamefont {Fu},\ and\ \citenamefont
  {Ju}}]{lu2023fractional}%
  \BibitemOpen
  \bibfield  {author} {\bibinfo {author} {\bibfnamefont {Z.}~\bibnamefont
  {Lu}}, \bibinfo {author} {\bibfnamefont {T.}~\bibnamefont {Han}}, \bibinfo
  {author} {\bibfnamefont {Y.}~\bibnamefont {Yao}}, \bibinfo {author}
  {\bibfnamefont {A.~P.}\ \bibnamefont {Reddy}}, \bibinfo {author}
  {\bibfnamefont {J.}~\bibnamefont {Yang}}, \bibinfo {author} {\bibfnamefont
  {J.}~\bibnamefont {Seo}}, \bibinfo {author} {\bibfnamefont {K.}~\bibnamefont
  {Watanabe}}, \bibinfo {author} {\bibfnamefont {T.}~\bibnamefont {Taniguchi}},
  \bibinfo {author} {\bibfnamefont {L.}~\bibnamefont {Fu}},\ and\ \bibinfo
  {author} {\bibfnamefont {L.}~\bibnamefont {Ju}},\ }\bibfield  {title}
  {\bibinfo {title} {Fractional quantum anomalous hall effect in a graphene
  moire superlattice},\ }\href@noop {} {\bibfield  {journal} {\bibinfo
  {journal} {arXiv preprint arXiv:2309.17436}\ } (\bibinfo {year}
  {2023})}\BibitemShut {NoStop}%
\bibitem [{\citenamefont {Han}\ \emph {et~al.}(2023{\natexlab{b}})\citenamefont
  {Han}, \citenamefont {Lu}, \citenamefont {Yao}, \citenamefont {Yang},
  \citenamefont {Seo}, \citenamefont {Yoon}, \citenamefont {Watanabe},
  \citenamefont {Taniguchi}, \citenamefont {Fu}, \citenamefont {Zhang} \emph
  {et~al.}}]{han2023large}%
  \BibitemOpen
  \bibfield  {author} {\bibinfo {author} {\bibfnamefont {T.}~\bibnamefont
  {Han}}, \bibinfo {author} {\bibfnamefont {Z.}~\bibnamefont {Lu}}, \bibinfo
  {author} {\bibfnamefont {Y.}~\bibnamefont {Yao}}, \bibinfo {author}
  {\bibfnamefont {J.}~\bibnamefont {Yang}}, \bibinfo {author} {\bibfnamefont
  {J.}~\bibnamefont {Seo}}, \bibinfo {author} {\bibfnamefont {C.}~\bibnamefont
  {Yoon}}, \bibinfo {author} {\bibfnamefont {K.}~\bibnamefont {Watanabe}},
  \bibinfo {author} {\bibfnamefont {T.}~\bibnamefont {Taniguchi}}, \bibinfo
  {author} {\bibfnamefont {L.}~\bibnamefont {Fu}}, \bibinfo {author}
  {\bibfnamefont {F.}~\bibnamefont {Zhang}}, \emph {et~al.},\ }\bibfield
  {title} {\bibinfo {title} {Large quantum anomalous hall effect in spin-orbit
  proximitized rhombohedral graphene},\ }\href@noop {} {\bibfield  {journal}
  {\bibinfo  {journal} {arXiv preprint arXiv:2310.17483}\ } (\bibinfo {year}
  {2023}{\natexlab{b}})}\BibitemShut {NoStop}%
\bibitem [{\citenamefont {Wu}\ \emph {et~al.}(2019)\citenamefont {Wu},
  \citenamefont {Lovorn}, \citenamefont {Tutuc}, \citenamefont {Martin},\ and\
  \citenamefont {MacDonald}}]{wu2019topological}%
  \BibitemOpen
  \bibfield  {author} {\bibinfo {author} {\bibfnamefont {F.}~\bibnamefont
  {Wu}}, \bibinfo {author} {\bibfnamefont {T.}~\bibnamefont {Lovorn}}, \bibinfo
  {author} {\bibfnamefont {E.}~\bibnamefont {Tutuc}}, \bibinfo {author}
  {\bibfnamefont {I.}~\bibnamefont {Martin}},\ and\ \bibinfo {author}
  {\bibfnamefont {A.~H.}\ \bibnamefont {MacDonald}},\ }\bibfield  {title}
  {\bibinfo {title} {Topological insulators in twisted transition metal
  dichalcogenide homobilayers},\ }\href
  {https://doi.org/10.1103/PhysRevLett.122.086402} {\bibfield  {journal}
  {\bibinfo  {journal} {Phys. Rev. Lett.}\ }\textbf {\bibinfo {volume} {122}},\
  \bibinfo {pages} {086402} (\bibinfo {year} {2019})}\BibitemShut {NoStop}%
\bibitem [{\citenamefont {Devakul}\ \emph {et~al.}(2021)\citenamefont
  {Devakul}, \citenamefont {Cr{\'e}pel}, \citenamefont {Zhang},\ and\
  \citenamefont {Fu}}]{devakul2021magic}%
  \BibitemOpen
  \bibfield  {author} {\bibinfo {author} {\bibfnamefont {T.}~\bibnamefont
  {Devakul}}, \bibinfo {author} {\bibfnamefont {V.}~\bibnamefont {Cr{\'e}pel}},
  \bibinfo {author} {\bibfnamefont {Y.}~\bibnamefont {Zhang}},\ and\ \bibinfo
  {author} {\bibfnamefont {L.}~\bibnamefont {Fu}},\ }\bibfield  {title}
  {\bibinfo {title} {Magic in twisted transition metal dichalcogenide
  bilayers},\ }\href@noop {} {\bibfield  {journal} {\bibinfo  {journal} {Nature
  communications}\ }\textbf {\bibinfo {volume} {12}},\ \bibinfo {pages} {6730}
  (\bibinfo {year} {2021})}\BibitemShut {NoStop}%
\bibitem [{\citenamefont {Cr{\'e}pel}\ and\ \citenamefont
  {Fu}(2023)}]{crepel2023anomalous}%
  \BibitemOpen
  \bibfield  {author} {\bibinfo {author} {\bibfnamefont {V.}~\bibnamefont
  {Cr{\'e}pel}}\ and\ \bibinfo {author} {\bibfnamefont {L.}~\bibnamefont
  {Fu}},\ }\bibfield  {title} {\bibinfo {title} {Anomalous hall metal and
  fractional chern insulator in twisted transition metal dichalcogenides},\
  }\href@noop {} {\bibfield  {journal} {\bibinfo  {journal} {Physical Review
  B}\ }\textbf {\bibinfo {volume} {107}},\ \bibinfo {pages} {L201109} (\bibinfo
  {year} {2023})}\BibitemShut {NoStop}%
\bibitem [{\citenamefont {Zeng}\ \emph {et~al.}(2024)\citenamefont {Zeng},
  \citenamefont {Wolf}, \citenamefont {Huang}, \citenamefont {Wei},
  \citenamefont {Ghorashi}, \citenamefont {MacDonald},\ and\ \citenamefont
  {Cano}}]{zeng2024gate}%
  \BibitemOpen
  \bibfield  {author} {\bibinfo {author} {\bibfnamefont {Y.}~\bibnamefont
  {Zeng}}, \bibinfo {author} {\bibfnamefont {T.~M.}\ \bibnamefont {Wolf}},
  \bibinfo {author} {\bibfnamefont {C.}~\bibnamefont {Huang}}, \bibinfo
  {author} {\bibfnamefont {N.}~\bibnamefont {Wei}}, \bibinfo {author}
  {\bibfnamefont {S.~A.~A.}\ \bibnamefont {Ghorashi}}, \bibinfo {author}
  {\bibfnamefont {A.~H.}\ \bibnamefont {MacDonald}},\ and\ \bibinfo {author}
  {\bibfnamefont {J.}~\bibnamefont {Cano}},\ }\bibfield  {title} {\bibinfo
  {title} {Gate-tunable topological phases in superlattice modulated bilayer
  graphene},\ }\href@noop {} {\bibfield  {journal} {\bibinfo  {journal} {arXiv
  preprint arXiv:2401.04321}\ } (\bibinfo {year} {2024})}\BibitemShut {NoStop}%
\bibitem [{\citenamefont {Tan}\ \emph {et~al.}(2024)\citenamefont {Tan},
  \citenamefont {Reddy}, \citenamefont {Fu},\ and\ \citenamefont
  {Devakul}}]{tan2024designing}%
  \BibitemOpen
  \bibfield  {author} {\bibinfo {author} {\bibfnamefont {T.}~\bibnamefont
  {Tan}}, \bibinfo {author} {\bibfnamefont {A.~P.}\ \bibnamefont {Reddy}},
  \bibinfo {author} {\bibfnamefont {L.}~\bibnamefont {Fu}},\ and\ \bibinfo
  {author} {\bibfnamefont {T.}~\bibnamefont {Devakul}},\ }\bibfield  {title}
  {\bibinfo {title} {Designing topology and fractionalization in narrow gap
  semiconductor films via electrostatic engineering},\ }\href@noop {}
  {\bibfield  {journal} {\bibinfo  {journal} {arXiv preprint arXiv:2402.03085}\
  } (\bibinfo {year} {2024})}\BibitemShut {NoStop}%
\bibitem [{\citenamefont {Su}\ \emph {et~al.}(2022)\citenamefont {Su},
  \citenamefont {Li}, \citenamefont {Zhang}, \citenamefont {Sun},\ and\
  \citenamefont {Lin}}]{su2022massive}%
  \BibitemOpen
  \bibfield  {author} {\bibinfo {author} {\bibfnamefont {Y.}~\bibnamefont
  {Su}}, \bibinfo {author} {\bibfnamefont {H.}~\bibnamefont {Li}}, \bibinfo
  {author} {\bibfnamefont {C.}~\bibnamefont {Zhang}}, \bibinfo {author}
  {\bibfnamefont {K.}~\bibnamefont {Sun}},\ and\ \bibinfo {author}
  {\bibfnamefont {S.-Z.}\ \bibnamefont {Lin}},\ }\bibfield  {title} {\bibinfo
  {title} {Massive dirac fermions in moir\'e superlattices: A route towards
  topological flat minibands and correlated topological insulators},\ }\href
  {https://doi.org/10.1103/PhysRevResearch.4.L032024} {\bibfield  {journal}
  {\bibinfo  {journal} {Phys. Rev. Res.}\ }\textbf {\bibinfo {volume} {4}},\
  \bibinfo {pages} {L032024} (\bibinfo {year} {2022})}\BibitemShut {NoStop}%
\bibitem [{\citenamefont {Cr\'epel}\ \emph {et~al.}(2023)\citenamefont
  {Cr\'epel}, \citenamefont {Dunbrack}, \citenamefont {Guerci}, \citenamefont
  {Bonini},\ and\ \citenamefont {Cano}}]{crepel2023chiraljen}%
  \BibitemOpen
  \bibfield  {author} {\bibinfo {author} {\bibfnamefont {V.}~\bibnamefont
  {Cr\'epel}}, \bibinfo {author} {\bibfnamefont {A.}~\bibnamefont {Dunbrack}},
  \bibinfo {author} {\bibfnamefont {D.}~\bibnamefont {Guerci}}, \bibinfo
  {author} {\bibfnamefont {J.}~\bibnamefont {Bonini}},\ and\ \bibinfo {author}
  {\bibfnamefont {J.}~\bibnamefont {Cano}},\ }\bibfield  {title} {\bibinfo
  {title} {Chiral model of twisted bilayer graphene realized in a monolayer},\
  }\href {https://doi.org/10.1103/PhysRevB.108.075126} {\bibfield  {journal}
  {\bibinfo  {journal} {Phys. Rev. B}\ }\textbf {\bibinfo {volume} {108}},\
  \bibinfo {pages} {075126} (\bibinfo {year} {2023})}\BibitemShut {NoStop}%
\bibitem [{\citenamefont {Halperin}\ \emph {et~al.}(1986)\citenamefont
  {Halperin}, \citenamefont {Te\ifmmode \check{s}\else
  \v{s}\fi{}anovi\ifmmode~\acute{c}\else \'{c}\fi{}},\ and\ \citenamefont
  {Axel}}]{halperin1986compatibility}%
  \BibitemOpen
  \bibfield  {author} {\bibinfo {author} {\bibfnamefont {B.~I.}\ \bibnamefont
  {Halperin}}, \bibinfo {author} {\bibfnamefont {Z.}~\bibnamefont {Te\ifmmode
  \check{s}\else \v{s}\fi{}anovi\ifmmode~\acute{c}\else \'{c}\fi{}}},\ and\
  \bibinfo {author} {\bibfnamefont {F.}~\bibnamefont {Axel}},\ }\bibfield
  {title} {\bibinfo {title} {Compatibility of crystalline order and the
  quantized hall effect},\ }\href {https://doi.org/10.1103/PhysRevLett.57.922}
  {\bibfield  {journal} {\bibinfo  {journal} {Phys. Rev. Lett.}\ }\textbf
  {\bibinfo {volume} {57}},\ \bibinfo {pages} {922} (\bibinfo {year}
  {1986})}\BibitemShut {NoStop}%
\bibitem [{\citenamefont {Te{\v{s}}anovi{\'c}}\ \emph
  {et~al.}(1989)\citenamefont {Te{\v{s}}anovi{\'c}}, \citenamefont {Axel},\
  and\ \citenamefont {Halperin}}]{tesanovic1989hall}%
  \BibitemOpen
  \bibfield  {author} {\bibinfo {author} {\bibfnamefont {Z.}~\bibnamefont
  {Te{\v{s}}anovi{\'c}}}, \bibinfo {author} {\bibfnamefont {F.}~\bibnamefont
  {Axel}},\ and\ \bibinfo {author} {\bibfnamefont {B.}~\bibnamefont
  {Halperin}},\ }\bibfield  {title} {\bibinfo {title} {{``Hall crystal'' versus
  Wigner crystal}},\ }\href {https://doi.org/10.1103/PhysRevB.39.8525}
  {\bibfield  {journal} {\bibinfo  {journal} {Phys. Rev. B}\ }\textbf {\bibinfo
  {volume} {39}},\ \bibinfo {pages} {8525} (\bibinfo {year}
  {1989})}\BibitemShut {NoStop}%
\bibitem [{\citenamefont {Dong}\ \emph
  {et~al.}(2023{\natexlab{a}})\citenamefont {Dong}, \citenamefont {Wang},
  \citenamefont {Wang}, \citenamefont {Soejima}, \citenamefont {Zaletel},
  \citenamefont {Vishwanath},\ and\ \citenamefont
  {Parker}}]{dong2023anomalous}%
  \BibitemOpen
  \bibfield  {author} {\bibinfo {author} {\bibfnamefont {J.}~\bibnamefont
  {Dong}}, \bibinfo {author} {\bibfnamefont {T.}~\bibnamefont {Wang}}, \bibinfo
  {author} {\bibfnamefont {T.}~\bibnamefont {Wang}}, \bibinfo {author}
  {\bibfnamefont {T.}~\bibnamefont {Soejima}}, \bibinfo {author} {\bibfnamefont
  {M.~P.}\ \bibnamefont {Zaletel}}, \bibinfo {author} {\bibfnamefont
  {A.}~\bibnamefont {Vishwanath}},\ and\ \bibinfo {author} {\bibfnamefont
  {D.~E.}\ \bibnamefont {Parker}},\ }\bibfield  {title} {\bibinfo {title}
  {{Anomalous Hall crystals in rhombohedral multilayer graphene I:
  Interaction-driven Chern bands and fractional quantum Hall states at zero
  magnetic field}},\ }\href@noop {} {\bibfield  {journal} {\bibinfo  {journal}
  {arXiv preprint arXiv:2311.05568}\ } (\bibinfo {year}
  {2023}{\natexlab{a}})}\BibitemShut {NoStop}%
\bibitem [{\citenamefont {Dong}\ \emph
  {et~al.}(2023{\natexlab{b}})\citenamefont {Dong}, \citenamefont {Patri},\
  and\ \citenamefont {Senthil}}]{dong2023theory}%
  \BibitemOpen
  \bibfield  {author} {\bibinfo {author} {\bibfnamefont {Z.}~\bibnamefont
  {Dong}}, \bibinfo {author} {\bibfnamefont {A.~S.}\ \bibnamefont {Patri}},\
  and\ \bibinfo {author} {\bibfnamefont {T.}~\bibnamefont {Senthil}},\
  }\bibfield  {title} {\bibinfo {title} {{Theory of fractional quantum
  anomalous Hall phases in pentalayer rhombohedral graphene moir\'e
  structures}},\ }\href@noop {} {\bibfield  {journal} {\bibinfo  {journal}
  {arXiv preprint arXiv:2311.03445}\ } (\bibinfo {year}
  {2023}{\natexlab{b}})}\BibitemShut {NoStop}%
\bibitem [{\citenamefont {Zhou}\ \emph {et~al.}(2023)\citenamefont {Zhou},
  \citenamefont {Yang},\ and\ \citenamefont {Zhang}}]{zhou2023fractional}%
  \BibitemOpen
  \bibfield  {author} {\bibinfo {author} {\bibfnamefont {B.}~\bibnamefont
  {Zhou}}, \bibinfo {author} {\bibfnamefont {H.}~\bibnamefont {Yang}},\ and\
  \bibinfo {author} {\bibfnamefont {Y.-H.}\ \bibnamefont {Zhang}},\ }\bibfield
  {title} {\bibinfo {title} {{Fractional quantum anomalous Hall effects in
  rhombohedral multilayer graphene in the moir\'eless limit and in Coulomb
  imprinted superlattice}},\ }\href@noop {} {\bibfield  {journal} {\bibinfo
  {journal} {arXiv preprint arXiv:2311.04217}\ } (\bibinfo {year}
  {2023})}\BibitemShut {NoStop}%
\bibitem [{\citenamefont {Guo}\ \emph {et~al.}(2023)\citenamefont {Guo},
  \citenamefont {Lu}, \citenamefont {Xie},\ and\ \citenamefont
  {Liu}}]{guo2023theory}%
  \BibitemOpen
  \bibfield  {author} {\bibinfo {author} {\bibfnamefont {Z.}~\bibnamefont
  {Guo}}, \bibinfo {author} {\bibfnamefont {X.}~\bibnamefont {Lu}}, \bibinfo
  {author} {\bibfnamefont {B.}~\bibnamefont {Xie}},\ and\ \bibinfo {author}
  {\bibfnamefont {J.}~\bibnamefont {Liu}},\ }\bibfield  {title} {\bibinfo
  {title} {{Theory of fractional Chern insulator states in pentalayer graphene
  moir\'e superlattice}},\ }\href@noop {} {\bibfield  {journal} {\bibinfo
  {journal} {arXiv preprint arXiv:2311.14368}\ } (\bibinfo {year}
  {2023})}\BibitemShut {NoStop}%
\bibitem [{\citenamefont {Kwan}\ \emph {et~al.}(2023)\citenamefont {Kwan},
  \citenamefont {Yu}, \citenamefont {Herzog-Arbeitman}, \citenamefont {Efetov},
  \citenamefont {Regnault},\ and\ \citenamefont {Bernevig}}]{kwan2023moire}%
  \BibitemOpen
  \bibfield  {author} {\bibinfo {author} {\bibfnamefont {Y.~H.}\ \bibnamefont
  {Kwan}}, \bibinfo {author} {\bibfnamefont {J.}~\bibnamefont {Yu}}, \bibinfo
  {author} {\bibfnamefont {J.}~\bibnamefont {Herzog-Arbeitman}}, \bibinfo
  {author} {\bibfnamefont {D.~K.}\ \bibnamefont {Efetov}}, \bibinfo {author}
  {\bibfnamefont {N.}~\bibnamefont {Regnault}},\ and\ \bibinfo {author}
  {\bibfnamefont {B.~A.}\ \bibnamefont {Bernevig}},\ }\bibfield  {title}
  {\bibinfo {title} {{Moir\'e Fractional Chern Insulators III: Hartree-Fock
  Phase Diagram, Magic Angle Regime for Chern Insulator States, the Role of the
  Moir\'e Potential and Goldstone Gaps in Rhombohedral Graphene
  Superlattices}},\ }\href@noop {} {\bibfield  {journal} {\bibinfo  {journal}
  {arXiv preprint arXiv:2312.11617}\ } (\bibinfo {year} {2023})}\BibitemShut
  {NoStop}%
\bibitem [{Note1()}]{Note1}%
  \BibitemOpen
  \bibinfo {note} {Note that $\protect \mathcal {T}$ is not the physical
  time-reversal operator, but rather an anti-unitary operator that acts like
  time-reversal within a single valley. The complete physical system contains
  another valley that is the physical time-reversal partner of the band we
  describe and transforms differently under the effective time-reversal
  $\protect \mathcal {T}$, but we assume it is at higher energy due to either
  explicit or spontaneous breaking of the physical time-reversal
  symmetry.}\BibitemShut {Stop}%
\bibitem [{Note2()}]{Note2}%
  \BibitemOpen
  \bibinfo {note} {Throughout this paper, by TRS we always refer to the
  effective time-reversal symmetry that acts within a single
  valley.}\BibitemShut {Stop}%
\bibitem [{\citenamefont {MacDonald}\ \emph {et~al.}(2012)\citenamefont
  {MacDonald}, \citenamefont {Jung},\ and\ \citenamefont
  {Zhang}}]{macdonald2012pseudospin}%
  \BibitemOpen
  \bibfield  {author} {\bibinfo {author} {\bibfnamefont {A.~H.}\ \bibnamefont
  {MacDonald}}, \bibinfo {author} {\bibfnamefont {J.}~\bibnamefont {Jung}},\
  and\ \bibinfo {author} {\bibfnamefont {F.}~\bibnamefont {Zhang}},\ }\bibfield
   {title} {\bibinfo {title} {Pseudospin order in monolayer, bilayer and
  double-layer graphene},\ }\href@noop {} {\bibfield  {journal} {\bibinfo
  {journal} {Physica Scripta}\ }\textbf {\bibinfo {volume} {2012}},\ \bibinfo
  {pages} {014012} (\bibinfo {year} {2012})}\BibitemShut {NoStop}%
\bibitem [{\citenamefont {Min}\ \emph {et~al.}(2008)\citenamefont {Min},
  \citenamefont {Borghi}, \citenamefont {Polini},\ and\ \citenamefont
  {MacDonald}}]{min2008pseudospin}%
  \BibitemOpen
  \bibfield  {author} {\bibinfo {author} {\bibfnamefont {H.}~\bibnamefont
  {Min}}, \bibinfo {author} {\bibfnamefont {G.}~\bibnamefont {Borghi}},
  \bibinfo {author} {\bibfnamefont {M.}~\bibnamefont {Polini}},\ and\ \bibinfo
  {author} {\bibfnamefont {A.~H.}\ \bibnamefont {MacDonald}},\ }\bibfield
  {title} {\bibinfo {title} {Pseudospin magnetism in graphene},\ }\href
  {https://doi.org/10.1103/PhysRevB.77.041407} {\bibfield  {journal} {\bibinfo
  {journal} {Phys. Rev. B}\ }\textbf {\bibinfo {volume} {77}},\ \bibinfo
  {pages} {041407} (\bibinfo {year} {2008})}\BibitemShut {NoStop}%
\bibitem [{Note3()}]{Note3}%
  \BibitemOpen
  \bibinfo {note} {Note that this is not a general result for any form factors.
  For example, for $N=1$ both $\protect \bm {h}^{\protect \rm HF}$ and $J\cdot
  n^{\protect \rm kin}$ favor $C=+1$. See Supplemental Material for results at
  other winding numbers.}\BibitemShut {Stop}%
\bibitem [{\citenamefont {Khrapak}(2020)}]{khrapak2020lindemann}%
  \BibitemOpen
  \bibfield  {author} {\bibinfo {author} {\bibfnamefont {S.~A.}\ \bibnamefont
  {Khrapak}},\ }\bibfield  {title} {\bibinfo {title} {Lindemann melting
  criterion in two dimensions},\ }\href
  {https://doi.org/10.1103/PhysRevResearch.2.012040} {\bibfield  {journal}
  {\bibinfo  {journal} {Phys. Rev. Res.}\ }\textbf {\bibinfo {volume} {2}},\
  \bibinfo {pages} {012040} (\bibinfo {year} {2020})}\BibitemShut {NoStop}%
\bibitem [{\citenamefont {Zheng}\ and\ \citenamefont
  {Earnshaw}(1998)}]{zheng1998lindemann}%
  \BibitemOpen
  \bibfield  {author} {\bibinfo {author} {\bibfnamefont {X.}~\bibnamefont
  {Zheng}}\ and\ \bibinfo {author} {\bibfnamefont {J.}~\bibnamefont
  {Earnshaw}},\ }\bibfield  {title} {\bibinfo {title} {{On the Lindemann
  criterion in 2D}},\ }\href@noop {} {\bibfield  {journal} {\bibinfo  {journal}
  {Europhysics Letters}\ }\textbf {\bibinfo {volume} {41}},\ \bibinfo {pages}
  {635} (\bibinfo {year} {1998})}\BibitemShut {NoStop}%
\bibitem [{\citenamefont {Bedanov}\ \emph {et~al.}(1985)\citenamefont
  {Bedanov}, \citenamefont {Gadiyak},\ and\ \citenamefont
  {Lozovik}}]{bedanov1985modified}%
  \BibitemOpen
  \bibfield  {author} {\bibinfo {author} {\bibfnamefont {V.}~\bibnamefont
  {Bedanov}}, \bibinfo {author} {\bibfnamefont {G.}~\bibnamefont {Gadiyak}},\
  and\ \bibinfo {author} {\bibfnamefont {Y.~E.}\ \bibnamefont {Lozovik}},\
  }\bibfield  {title} {\bibinfo {title} {On a modified lindemann-like criterion
  for 2d melting},\ }\href@noop {} {\bibfield  {journal} {\bibinfo  {journal}
  {Physics Letters A}\ }\textbf {\bibinfo {volume} {109}},\ \bibinfo {pages}
  {289} (\bibinfo {year} {1985})}\BibitemShut {NoStop}%
\bibitem [{\citenamefont {Goldoni}\ and\ \citenamefont
  {Peeters}(1996)}]{goldoni1996stability}%
  \BibitemOpen
  \bibfield  {author} {\bibinfo {author} {\bibfnamefont {G.}~\bibnamefont
  {Goldoni}}\ and\ \bibinfo {author} {\bibfnamefont {F.~M.}\ \bibnamefont
  {Peeters}},\ }\bibfield  {title} {\bibinfo {title} {Stability, dynamical
  properties, and melting of a classical bilayer wigner crystal},\ }\href
  {https://doi.org/10.1103/PhysRevB.53.4591} {\bibfield  {journal} {\bibinfo
  {journal} {Phys. Rev. B}\ }\textbf {\bibinfo {volume} {53}},\ \bibinfo
  {pages} {4591} (\bibinfo {year} {1996})}\BibitemShut {NoStop}%
\bibitem [{\citenamefont {Wang}\ \emph {et~al.}(2021)\citenamefont {Wang},
  \citenamefont {Cano}, \citenamefont {Millis}, \citenamefont {Liu},\ and\
  \citenamefont {Yang}}]{wang2021exact}%
  \BibitemOpen
  \bibfield  {author} {\bibinfo {author} {\bibfnamefont {J.}~\bibnamefont
  {Wang}}, \bibinfo {author} {\bibfnamefont {J.}~\bibnamefont {Cano}}, \bibinfo
  {author} {\bibfnamefont {A.~J.}\ \bibnamefont {Millis}}, \bibinfo {author}
  {\bibfnamefont {Z.}~\bibnamefont {Liu}},\ and\ \bibinfo {author}
  {\bibfnamefont {B.}~\bibnamefont {Yang}},\ }\bibfield  {title} {\bibinfo
  {title} {Exact landau level description of geometry and interaction in a
  flatband},\ }\href@noop {} {\bibfield  {journal} {\bibinfo  {journal}
  {Physical review letters}\ }\textbf {\bibinfo {volume} {127}},\ \bibinfo
  {pages} {246403} (\bibinfo {year} {2021})}\BibitemShut {NoStop}%
\bibitem [{\citenamefont {Estienne}\ \emph {et~al.}(2023)\citenamefont
  {Estienne}, \citenamefont {Regnault},\ and\ \citenamefont
  {Cr\'epel}}]{estienne2023ideal}%
  \BibitemOpen
  \bibfield  {author} {\bibinfo {author} {\bibfnamefont {B.}~\bibnamefont
  {Estienne}}, \bibinfo {author} {\bibfnamefont {N.}~\bibnamefont {Regnault}},\
  and\ \bibinfo {author} {\bibfnamefont {V.}~\bibnamefont {Cr\'epel}},\
  }\bibfield  {title} {\bibinfo {title} {Ideal chern bands as landau levels in
  curved space},\ }\href {https://doi.org/10.1103/PhysRevResearch.5.L032048}
  {\bibfield  {journal} {\bibinfo  {journal} {Phys. Rev. Res.}\ }\textbf
  {\bibinfo {volume} {5}},\ \bibinfo {pages} {L032048} (\bibinfo {year}
  {2023})}\BibitemShut {NoStop}%
\bibitem [{\citenamefont {Cr{\'e}pel}\ \emph {et~al.}(2023)\citenamefont
  {Cr{\'e}pel}, \citenamefont {Regnault},\ and\ \citenamefont
  {Queiroz}}]{crepel2023chiral}%
  \BibitemOpen
  \bibfield  {author} {\bibinfo {author} {\bibfnamefont {V.}~\bibnamefont
  {Cr{\'e}pel}}, \bibinfo {author} {\bibfnamefont {N.}~\bibnamefont
  {Regnault}},\ and\ \bibinfo {author} {\bibfnamefont {R.}~\bibnamefont
  {Queiroz}},\ }\bibfield  {title} {\bibinfo {title} {The chiral limits of
  moir$\backslash$'e semiconductors: origin of flat bands and topology in
  twisted transition metal dichalcogenides homobilayers},\ }\href@noop {}
  {\bibfield  {journal} {\bibinfo  {journal} {arXiv preprint arXiv:2305.10477}\
  } (\bibinfo {year} {2023})}\BibitemShut {NoStop}%
\bibitem [{\citenamefont {Marzari}\ and\ \citenamefont
  {Vanderbilt}(1997)}]{marzari1997maximally}%
  \BibitemOpen
  \bibfield  {author} {\bibinfo {author} {\bibfnamefont {N.}~\bibnamefont
  {Marzari}}\ and\ \bibinfo {author} {\bibfnamefont {D.}~\bibnamefont
  {Vanderbilt}},\ }\bibfield  {title} {\bibinfo {title} {Maximally localized
  generalized wannier functions for composite energy bands},\ }\href
  {https://doi.org/10.1103/PhysRevB.56.12847} {\bibfield  {journal} {\bibinfo
  {journal} {Phys. Rev. B}\ }\textbf {\bibinfo {volume} {56}},\ \bibinfo
  {pages} {12847} (\bibinfo {year} {1997})}\BibitemShut {NoStop}%
\bibitem [{\citenamefont {Souza}\ \emph {et~al.}(2001)\citenamefont {Souza},
  \citenamefont {Marzari},\ and\ \citenamefont
  {Vanderbilt}}]{souza2001maximally}%
  \BibitemOpen
  \bibfield  {author} {\bibinfo {author} {\bibfnamefont {I.}~\bibnamefont
  {Souza}}, \bibinfo {author} {\bibfnamefont {N.}~\bibnamefont {Marzari}},\
  and\ \bibinfo {author} {\bibfnamefont {D.}~\bibnamefont {Vanderbilt}},\
  }\bibfield  {title} {\bibinfo {title} {Maximally localized wannier functions
  for entangled energy bands},\ }\href
  {https://doi.org/10.1103/PhysRevB.65.035109} {\bibfield  {journal} {\bibinfo
  {journal} {Phys. Rev. B}\ }\textbf {\bibinfo {volume} {65}},\ \bibinfo
  {pages} {035109} (\bibinfo {year} {2001})}\BibitemShut {NoStop}%
\bibitem [{\citenamefont {Marzari}\ \emph {et~al.}(2012)\citenamefont
  {Marzari}, \citenamefont {Mostofi}, \citenamefont {Yates}, \citenamefont
  {Souza},\ and\ \citenamefont {Vanderbilt}}]{marzari2012maximally}%
  \BibitemOpen
  \bibfield  {author} {\bibinfo {author} {\bibfnamefont {N.}~\bibnamefont
  {Marzari}}, \bibinfo {author} {\bibfnamefont {A.~A.}\ \bibnamefont
  {Mostofi}}, \bibinfo {author} {\bibfnamefont {J.~R.}\ \bibnamefont {Yates}},
  \bibinfo {author} {\bibfnamefont {I.}~\bibnamefont {Souza}},\ and\ \bibinfo
  {author} {\bibfnamefont {D.}~\bibnamefont {Vanderbilt}},\ }\bibfield  {title}
  {\bibinfo {title} {Maximally localized wannier functions: Theory and
  applications},\ }\href {https://doi.org/10.1103/RevModPhys.84.1419}
  {\bibfield  {journal} {\bibinfo  {journal} {Rev. Mod. Phys.}\ }\textbf
  {\bibinfo {volume} {84}},\ \bibinfo {pages} {1419} (\bibinfo {year}
  {2012})}\BibitemShut {NoStop}%
\end{thebibliography}%


\onecolumngrid
\newpage
\makeatletter 

\begin{center}
\textbf{\large Supplemental material for ``\@title ''} \\[10pt]
\end{center}
\vspace{20pt}

\setcounter{figure}{0}
\setcounter{section}{0}
\setcounter{equation}{0}

\renewcommand{\thefigure}{S\@arabic\c@figure}
\renewcommand{\theequation}{S\@arabic\c@equation}
\makeatother



\section{Pseudospin model}

In this section we derive the pseudospin model by calculating the energy expectation value of the Slater determinant
\begin{equation}
\ket{\Psi} = \prod_{\bm k \in {\rm mBZ}} \left(\cos\frac{\theta_{\bm k}}{2} a_{\bm k}^{\dagger} + e^{i\phi_{\bm k}} \sin\frac{\theta_{\bm k}}{2} b_{\bm k}^{\dagger} \right) \ket{0}.
\end{equation}
Here $a_{\bm k}^{\dagger} = \sum_{\bm g} A_{\bm k + \bm g} c_{\bm k + \bm g}^{\dagger}$ and $b_{\bm k}^{\dagger} = \sum_{\bm g} B_{\bm k + \bm g} c_{\bm k + \bm g}^{\dagger}$ are sublattice basis states that satisfy the orthonormality conditions
\begin{equation}
\sum_{\bm g} |A_{\bm k + \bm g}|^2 = \sum_{\bm g} |B_{\bm k + \bm g}|^2 = 1, \quad \sum_{\bm g} A_{\bm k + \bm g}^* B_{\bm k + \bm g} = 0.
\end{equation}
In practice the sublattice basis states are obtained by solving the problem of an electron moving in a honeycomb lattice potential in the plane wave basis and then Wannierizing the lowest two bands that touch at the Dirac points; see Sec.~\ref{sec:wannierize} for details. The system Hamiltonian is
\begin{align}
H &= H_{\rm kin} + H_{\rm SL} + H_{\rm int}, \\
H_{\rm kin} &= \sum_{\bm p} E_{\bm p} c_{\bm p}^{\dagger} c_{\bm p}, \\
H_{\rm SL} &= \sum_{\bm p \bm g} U_{\bm g} \Lambda_{\bm p + \bm g, \bm p} c_{\bm p + \bm g}^{\dagger} c_{\bm p}, \\
H_{\rm int} &= \frac{1}{2A} \sum_{\bm p \bm p' \bm q} V_{\bm q} \Lambda_{\bm p + \bm q, \bm p} \Lambda_{\bm p', \bm p' + \bm q} c_{\bm p + \bm q}^{\dagger} c_{\bm p'}^{\dagger} c_{\bm p' + \bm q} c_{\bm p},
\end{align}
where for completeness and pedagogical purpose we include a superlattice potential part $H_{\rm SL}$ that is not present in the Wigner crystal problem. The mean-field energy functional is obtained by calculating the energy expectation value
\begin{equation}
E_{\rm MF}[\theta,\phi] \equiv \braket{\Psi | H | \Psi} = E_{\rm kin} + E_{\rm SL} + E_{\rm H} + E_{\rm F}.
\end{equation}
The kinetic energy is
\begin{equation}
E_{\rm kin} = \braket{\Psi | H_{\rm kin} | \Psi} = \frac 12 \sum_{\bm k \bm g} E_{\bm k + \bm g} \big[ A^* A + B^* B + (A^* A - B^* B) \cos\theta_{\bm k} + (A^* B e^{i\phi_{\bm k}} + B^* A e^{-i\phi_{\bm k}}) \sin\theta_{\bm k} \big]_{(\bm k + \bm g, \bm k + \bm g)},
\end{equation}
where for notational simplicity we collectively write the momentum labels for $A$ and $B$, which are identical for all terms inside the square bracket, outside the bracket. The superlattice potential energy is
\begin{equation}
E_{\rm SL} = \braket{\Psi | H_{\rm SL} | \Psi} = \frac 12 \sum_{\bm k \bm g \bm g'} U_{\bm g'} \big[ \Lambda^{AA} + \Lambda^{BB} + (\Lambda^{AA}-\Lambda^{BB}) \cos\theta_{\bm k} + (\Lambda^{AB} e^{i\phi_{\bm k}} + \Lambda^{BA} e^{-i\phi_{\bm k}}) \sin\theta_{\bm k} \big]_{(\bm k + \bm g + \bm g', \bm k + \bm g)},
\end{equation}
where we introduced the shorthand notation $\Lambda^{AA}_{\bm p', \bm p} = A_{\bm p'}^* \Lambda_{\bm p', \bm p} A_{\bm p}$, etc. The interaction energy $\braket{\Psi | H_{\rm int} | \Psi}$ consists of the Hartree term
\begin{equation}
\begin{split}
E_{\rm H} = \frac{1}{8A} \sum_{\bm g''} &V_{\bm g''} \sum_{\bm k \bm g} \big[ \Lambda^{AA} + \Lambda^{BB} + (\Lambda^{AA}-\Lambda^{BB}) \cos\theta_{\bm k} + (\Lambda^{AB} e^{i\phi_{\bm k}} + \Lambda^{BA} e^{-i\phi_{\bm k}}) \sin\theta_{\bm k} \big]_{(\bm k + \bm g + \bm g'', \bm k + \bm g)} \\
&\times \sum_{\bm k' \bm g'} \big[ \Lambda^{AA} + \Lambda^{BB} + (\Lambda^{AA}-\Lambda^{BB}) \cos\theta_{\bm k'} + (\Lambda^{AB} e^{i\phi_{\bm k'}} + \Lambda^{BA} e^{-i\phi_{\bm k'}}) \sin\theta_{\bm k'} \big]_{(\bm k' + \bm g', \bm k' + \bm g' + \bm g'')}
\end{split}
\end{equation}
and the Fock term
\begin{equation}
\begin{split}
E_{\rm F} = -\frac{1}{8A} &\sum_{\substack{\bm k \bm k' \\ \bm g \bm g' \bm g''}} V_{\bm k' - \bm k + \bm g''} \Lambda_{\bm k' + \bm g + \bm g'', \bm k + \bm g} \Lambda_{\bm k + \bm g', \bm k' + \bm g' + \bm g''} \\
&\times \big[ A^* A + B^* B + (A^* A - B^* B) \cos\theta_{\bm k} + (A^* B e^{i\phi_{\bm k}} + B^* A e^{-i\phi_{\bm k}}) \sin\theta_{\bm k} \big]_{(\bm k + \bm g', \bm k + \bm g)} \\
&\times \big[ A^* A + B^* B + (A^* A - B^* B) \cos\theta_{\bm k'} + (A^* B e^{i\phi_{\bm k'}} + B^* A e^{-i\phi_{\bm k'}}) \sin\theta_{\bm k'} \big]_{(\bm k' + \bm g + \bm g'', \bm k' + \bm g' + \bm g'')}.
\end{split}
\end{equation}
Rewriting the energy functional in terms of sublattice pseudospin vectors
\begin{equation}
\bm{n_k} = (\sin\theta_{\bm k} \cos\phi_{\bm k}, \sin\theta_{\bm k} \sin\phi_{\bm k}, \cos\theta_{\bm k}),
\end{equation}
the system is effectively described by a spin model in momentum space:
\begin{equation} \label{eq:E_MF_SM}
E_{\rm MF}[\theta, \phi] = E_0 -\sum_{\bm k} \bm{h_k} \cdot \bm{n_k} - \frac{1}{2} \sum_{\alpha\beta} \sum_{\bm k \bm k'} J_{\bm k \bm k'}^{\alpha\beta} n_{\bm k}^{\alpha} n_{\bm k'}^{\beta},
\end{equation}
where $E_0$ is a constant energy independent of pseudospin orientations.

\subsection{Pseudospin Zeeman field}

The first-order coefficients in Eq.~\eqref{eq:E_MF_SM} act as an effective Zeeman field on sublattice pseudospins. The expressions are
\begin{align}
h_{\bm k}^x = &-\frac 12 \sum_{\bm g} E_{\bm k + \bm g} (A^* B + B^* A)_{(\bm k + \bm g, \bm k + \bm g)} - \frac 12 \sum_{\bm g \bm g'} U_{\bm g'} (\Lambda^{AB} + \Lambda^{BA})_{(\bm k + \bm g + \bm g', \bm k + \bm g)} \nonumber \\
&-\frac{1}{4A} \sum_{\bm k' \bm g''} V_{\bm g''} \sum_{\bm g'} (\Lambda^{AA} + \Lambda^{BB})_{(\bm k' + \bm g', \bm k' + \bm g' + \bm g'')} \sum_{\bm g} (\Lambda^{AB} + \Lambda^{BA})_{(\bm k + \bm g + \bm g'', \bm k + \bm g)} \nonumber \\
&+\frac{1}{4A} \sum_{\bm k' \bm g''} V_{\bm k' - \bm k + \bm g''} \sum_{\bm g \bm g'} (\Lambda^{AB} \Lambda^{AA} + \Lambda^{AA} \Lambda^{BA} + \Lambda^{BB} \Lambda^{AB} + \Lambda^{BA} \Lambda^{BB})_{(\bm k' + \bm g + \bm g'', \bm k + \bm g); (\bm k + \bm g', \bm k' + \bm g' + \bm g'')}, \\
h_{\bm k}^y = &-\frac i2 \sum_{\bm g} E_{\bm k + \bm g} (A^* B - B^* A)_{(\bm k + \bm g, \bm k + \bm g)} - \frac i2 \sum_{\bm g \bm g'} U_{\bm g'} (\Lambda^{AB} - \Lambda^{BA})_{(\bm k + \bm g + \bm g', \bm k + \bm g)} \nonumber \\
&-\frac{i}{4A} \sum_{\bm k' \bm g''} V_{\bm g''} \sum_{\bm g'} (\Lambda^{AA} + \Lambda^{BB})_{(\bm k' + \bm g', \bm k' + \bm g' + \bm g'')} \sum_{\bm g} (\Lambda^{AB} - \Lambda^{BA})_{(\bm k + \bm g + \bm g'', \bm k + \bm g)} \nonumber \\
&+\frac{i}{4A} \sum_{\bm k' \bm g''} V_{\bm k' - \bm k + \bm g''} \sum_{\bm g \bm g'} (\Lambda^{AB} \Lambda^{AA} - \Lambda^{AA} \Lambda^{BA} + \Lambda^{BB} \Lambda^{AB} - \Lambda^{BA} \Lambda^{BB})_{(\bm k' + \bm g + \bm g'', \bm k + \bm g); (\bm k + \bm g', \bm k' + \bm g' + \bm g'')}, \\
h_{\bm k}^z = &-\frac 12 \sum_{\bm g} E_{\bm k + \bm g} (A^* A - B^* B)_{(\bm k + \bm g, \bm k + \bm g)} - \frac 12 \sum_{\bm g \bm g'} U_{\bm g'} (\Lambda^{AA} - \Lambda^{BB})_{(\bm k + \bm g + \bm g', \bm k + \bm g)} \nonumber \\
&-\frac{1}{4A} \sum_{\bm k' \bm g''} V_{\bm g''} \sum_{\bm g'} (\Lambda^{AA} + \Lambda^{BB})_{(\bm k' + \bm g', \bm k' + \bm g' + \bm g'')} \sum_{\bm g} (\Lambda^{AA} - \Lambda^{BB})_{(\bm k + \bm g + \bm g'', \bm k + \bm g)} \nonumber \\
&+\frac{1}{4A} \sum_{\bm k' \bm g''} V_{\bm k' - \bm k + \bm g''} \sum_{\bm g \bm g'} (\Lambda^{AA} \Lambda^{AA} - \Lambda^{BB} \Lambda^{BB} - \Lambda^{AB} \Lambda^{BA} + \Lambda^{BA} \Lambda^{AB})_{(\bm k' + \bm g + \bm g'', \bm k + \bm g); (\bm k + \bm g', \bm k' + \bm g' + \bm g'')}.
\end{align}
Each expression above contains four terms that respectively come from the kinetic energy, superlattice potential energy, Hartree energy, and Fock energy.

\subsection{Pseudospin coupling coefficients}

The second-order terms in Eq.~\eqref{eq:E_MF_SM} represent coupling between pseudospins. The dominant terms are those in which $\alpha=\beta$:
\begingroup
\allowdisplaybreaks
\begin{align}
J_{\bm k \bm k'}^{xx} = &J_{\bm k' \bm k}^{xx} = -\frac{1}{4A} \sum_{\bm g''} V_{\bm g''} \sum_{\bm g} (\Lambda^{AB} + \Lambda^{BA})_{(\bm k + \bm g + \bm g'', \bm k + \bm g)} \sum_{\bm g'} (\Lambda^{AB} + \Lambda^{BA})_{(\bm k' + \bm g', \bm k' + \bm g' + \bm g'')} \nonumber \\
&+\frac{1}{4A} \sum_{\bm g''} V_{\bm k' - \bm k + \bm g''} \sum_{\bm g \bm g'} (\Lambda^{AA} \Lambda^{BB} + \Lambda^{BB} \Lambda^{AA} + \Lambda^{AB} \Lambda^{AB} + \Lambda^{BA} \Lambda^{BA})_{(\bm k' + \bm g + \bm g'', \bm k + \bm g); (\bm k + \bm g', \bm k' + \bm g' + \bm g'')}, \\
J_{\bm k \bm k'}^{yy} = &J_{\bm k' \bm k}^{yy} = \frac{1}{4A} \sum_{\bm g''} V_{\bm g''} \sum_{\bm g} (\Lambda^{AB} - \Lambda^{BA})_{(\bm k + \bm g + \bm g'', \bm k + \bm g)} \sum_{\bm g'} (\Lambda^{AB} - \Lambda^{BA})_{(\bm k' + \bm g', \bm k' + \bm g' + \bm g'')} \nonumber \\
&+\frac{1}{4A} \sum_{\bm g''} V_{\bm k' - \bm k + \bm g''} \sum_{\bm g \bm g'} (\Lambda^{AA} \Lambda^{BB} + \Lambda^{BB} \Lambda^{AA} - \Lambda^{AB} \Lambda^{AB} - \Lambda^{BA} \Lambda^{BA})_{(\bm k' + \bm g + \bm g'', \bm k + \bm g); (\bm k + \bm g', \bm k' + \bm g' + \bm g'')}, \\
J_{\bm k \bm k'}^{zz} = &J_{\bm k' \bm k}^{zz} = -\frac{1}{4A} \sum_{\bm g''} V_{\bm g''} \sum_{\bm g} (\Lambda^{AA} - \Lambda^{BB})_{(\bm k + \bm g + \bm g'', \bm k + \bm g)} \sum_{\bm g'} (\Lambda^{AA} - \Lambda^{BB})_{(\bm k' + \bm g', \bm k' + \bm g' + \bm g'')} \nonumber \\
&+\frac{1}{4A} \sum_{\bm g''} V_{\bm k' - \bm k + \bm g''} \sum_{\bm g \bm g'} (\Lambda^{AA} \Lambda^{AA} + \Lambda^{BB} \Lambda^{BB} - \Lambda^{AB} \Lambda^{BA} - \Lambda^{BA} \Lambda^{AB})_{(\bm k' + \bm g + \bm g'', \bm k + \bm g); (\bm k + \bm g', \bm k' + \bm g' + \bm g'')}.
\end{align}
\endgroup
These terms represent ferromagnetic coupling between pseudospins. Coupling between different components $\alpha\ne\beta$ is also generically nonvanishing:
\begin{align}
J_{\bm k \bm k'}^{xy} = &J_{\bm k' \bm k}^{yx} = -\frac{i}{4A} \sum_{\bm g''} V_{\bm g''} \sum_{\bm g} (\Lambda^{AB} + \Lambda^{BA})_{(\bm k + \bm g + \bm g'', \bm k + \bm g)} \sum_{\bm g'} (\Lambda^{AB} - \Lambda^{BA})_{(\bm k' + \bm g', \bm k' + \bm g' + \bm g'')} \nonumber \\
&+\frac{i}{4A} \sum_{\bm g''} V_{\bm k' - \bm k + \bm g''} \sum_{\bm g \bm g'} (\Lambda^{AA} \Lambda^{BB} - \Lambda^{BB} \Lambda^{AA} + \Lambda^{AB} \Lambda^{AB} - \Lambda^{BA} \Lambda^{BA})_{(\bm k' + \bm g + \bm g'', \bm k + \bm g); (\bm k + \bm g', \bm k' + \bm g' + \bm g'')}, \\
J_{\bm k \bm k'}^{zx} = &J_{\bm k' \bm k}^{xz} = -\frac{1}{4A} \sum_{\bm g''} V_{\bm g''} \sum_{\bm g} (\Lambda^{AA} - \Lambda^{BB})_{(\bm k + \bm g + \bm g'', \bm k + \bm g)} \sum_{\bm g'} (\Lambda^{AB} - \Lambda^{BA})_{(\bm k' + \bm g', \bm k' + \bm g' + \bm g'')} \nonumber \\
&+\frac{1}{4A} \sum_{\bm g''} V_{\bm k' - \bm k + \bm g''} \sum_{\bm g \bm g'} (\Lambda^{AA} \Lambda^{AB} + \Lambda^{BA} \Lambda^{AA} - \Lambda^{AB} \Lambda^{BB} - \Lambda^{BB} \Lambda^{BA})_{(\bm k' + \bm g + \bm g'', \bm k + \bm g); (\bm k + \bm g', \bm k' + \bm g' + \bm g'')}, \\
J_{\bm k \bm k'}^{zy} = &J_{\bm k' \bm k}^{yz} = -\frac{i}{4A} \sum_{\bm g''} V_{\bm g''} \sum_{\bm g} (\Lambda^{AA} - \Lambda^{BB})_{(\bm k + \bm g + \bm g'', \bm k + \bm g)} \sum_{\bm g'} (\Lambda^{AB} - \Lambda^{BA})_{(\bm k' + \bm g', \bm k' + \bm g' + \bm g'')} \nonumber \\
&+\frac{i}{4A} \sum_{\bm g''} V_{\bm k' - \bm k + \bm g''} \sum_{\bm g \bm g'} (\Lambda^{AA} \Lambda^{AB} - \Lambda^{BA} \Lambda^{AA} - \Lambda^{AB} \Lambda^{BB} + \Lambda^{BB} \Lambda^{BA})_{(\bm k' + \bm g + \bm g'', \bm k + \bm g); (\bm k + \bm g', \bm k' + \bm g' + \bm g'')}.
\end{align}

In the limit of small $|\bm k - \bm k'|$, the coupling coefficients are dominated by the exchange terms with $\bm g''=0$, and we have approximately $\sum_{\bm g} \Lambda^{AA}_{\bm k' + \bm g, \bm k + \bm g} \approx \sum_{\bm g} \Lambda^{BB}_{\bm k' + \bm g, \bm k + \bm g} \approx 1$ and $\sum_{\bm g} \Lambda^{AB}_{\bm k' + \bm g, \bm k + \bm g} \approx \sum_{\bm g} \Lambda^{BA}_{\bm k' + \bm g, \bm k + \bm g} \approx 0$. With this approximation the dominant coupling coefficients are
\begin{equation}
J_{\bm k \bm k'}^{xx} \approx J_{\bm k \bm k'}^{yy} \approx J_{\bm k \bm k'}^{zz} \approx \frac{1}{2A} V_{\bm k' - \bm k},
\end{equation}
and all other coupling coefficients are negligibly small. At the next-order expansion that distinguishes $\Lambda^{AA}$ and $\Lambda^{BB}$ but still neglects $\Lambda^{AB}$ and $\Lambda^{BA}$ (assuming localized sublattice basis states), $J^{zz}$ is slightly larger than $J^{xx}$ and $J^{yy}$. Physically this means that electrons tend to form sublattice polarized states rather than inter-sublattice dimers in the strong interaction limit.

\subsection{Symmetry analysis}

Symmetries of the system impose constraints on the $h$ and $J$ coefficients. We consider the following symmetries:
\begin{itemize}
    \item $C_6$ rotational symmetry of energy dispersion, interaction potential, and form factors: $E_{C_6 \bm p} = E_{\bm p}$, $V_{C_6 \bm q} = V_{\bm q}$, $\Lambda_{C_6 \bm p', C_6 \bm p} = \Lambda_{\bm p', \bm p}$. The superlattice potential is $C_6$-symmetric when $U_{\bm g} \in \mathbb{R}$.
    \item Two sublattice bases are related by $C_2$ around honeycomb center: $A_{\bm p} = \varphi_{\bm p} e^{-i\bm p \cdot \bm a_0}$, $B_{\bm p} = \varphi_{-\bm p} e^{-i\bm p \cdot \bm b_0}$ where $\bm a_0 = -\bm b_0 = (a/\sqrt{3},0)$. Each basis has $C_3$ and time-reversal symmetry: $\varphi_{C_3 \bm p} = \varphi_{\bm p}$, $\varphi_{-\bm p} = \varphi_{\bm p}^*$.
    \item When time-reversal symmetry (TRS) is present, $\Lambda_{-\bm p', -\bm p} = \Lambda_{\bm p', \bm p}^*$. Combined with $C_2$ this implies $\Lambda_{\bm p', \bm p} \in \mathbb{R}$.
\end{itemize}
We show the following consequences of these symmetries:
\begin{itemize}
    \item $C_3$ symmetry implies that the in-plane part of $\bm{h_k}$ forms vortices around $\bm K = (2\pi/\sqrt{3}a, 2\pi/3a)$ and $\bm K' = -\bm K$.
    \item $C_2$ symmetry implies $h_{-\bm k}^x = h_{\bm k}^x$, $h_{-\bm k}^y = -h_{\bm k}^y$, and $h_{-\bm k}^z = -h_{\bm k}^z$.
    \item If TRS is intact, $h_{-\bm k}^x = h_{\bm k}^x$, $h_{-\bm k}^y = -h_{\bm k}^y$, and $h_{-\bm k}^z = h_{\bm k}^z$. Combined with $C_2$ symmetry this implies vanishing of $h^z$, $J^{zx}$, and $J^{zy}$.
\end{itemize}
All these results can be verified by using the explicit expressions of $h$ and $J$ in the last section, but here we prove these results by examining the symmetry operations on sublattice pseudospins.

$C_3$ does not swap sublattices but rotates the momentum labels. More precisely,
\begin{equation}
C_3 a_{\bm k}^{\dagger} C_3^{-1} = \sum_{\bm g} \varphi_{\bm k + \bm g} e^{-i(\bm k + \bm g) \cdot \bm a_0} c_{C_3(\bm k + \bm g)}^{\dagger} = \sum_{\bm g} \varphi_{C_3 (\bm k + \bm g)} e^{-iC_3 (\bm k + \bm g) \cdot C_3 \bm a_0} c_{C_3(\bm k + \bm g)}^{\dagger} = e^{-i\bm k \cdot (\bm a_0 - C_3^{-1} \bm a_0)} a_{C_3 \bm k}^{\dagger},
\end{equation}
and similarly $C_3 b_{\bm k}^{\dagger} = e^{-i\bm k \cdot (\bm b_0 - C_3^{-1} \bm b_0)} b_{C_3 \bm k}^{\dagger}$. Due to the different phase factors on two sublattices, $C_3$ also rotates pseudospins around the $z$-axis:
\begin{equation}
\sum_{\bm k} \bm{h_k} \cdot \bm{n_k} \xrightarrow{C_3} \sum_{\bm k} \bm h_{C_3 \bm k} \cdot R_z \left(\bm k \cdot (1-C_3^{-1})(\bm a_0 - \bm b_0) \right) \bm n_{\bm k},
\end{equation}
where the operator $R_z(\theta)$ represents rotation by angle $\theta$ around the $z$-axis. $C_3$ invariance of the model requires that
\begin{equation}
\bm h_{C_3 \bm k} = R_z \left(\bm k \cdot (1-C_3^{-1})(\bm a_0 - \bm b_0) \right) \bm{h_k}.
\end{equation}
For $\bm k = \pm\bm K + \delta \bm k$ with $|\delta \bm k| \ll |\bm K|$, since $\pm\bm K$ is $C_3$-invariant, we have
\begin{equation}
\bm h_{\pm\bm K + C_3 \delta \bm k} \approx R_z(\pm 2\pi/3) \bm h_{\pm\bm K + \delta \bm k}.
\end{equation}
When $\delta\bm k = 0$ we have the equality $\bm h_{\pm\bm K} = R_z(\pm 2\pi/3) \bm h_{\pm\bm K}$ and hence $\bm h_{\pm\bm K} = 0$. It follows that $\bm{h_k}$ forms vortices of opposite chirality around $K$ and $K'$.

$C_2$ rotation inverts momenta as well as sublattices:
\begin{equation}
C_2 a_{\bm k}^{\dagger} C_2^{-1} = b_{-\bm k}^{\dagger}, \quad C_2 b_{\bm k}^{\dagger} C_2^{-1} = a_{-\bm k}^{\dagger}.
\end{equation}
In terms of pseudospins, $C_2$ inverts the $y,z$ components but not $x$ component:
\begin{equation}
(n_{\bm k}^x, n_{\bm k}^y, n_{\bm k}^z) \xrightarrow{C_2} (n_{-\bm k}^x, -n_{-\bm k}^y, -n_{-\bm k}^z).
\end{equation}
$C_2$-invariance of the model implies
\begin{equation}
h_{-\bm k}^x = h_{\bm k}^x, \quad h_{-\bm k}^y = -h_{\bm k}^y, \quad h_{-\bm k}^z = -h_{\bm k}^z.
\end{equation}

Time reversal $\mathcal{T}$ inverts momenta but keeps sublattices unchanged:
\begin{equation}
\mathcal{T} a_{\bm k}^{\dagger} \mathcal{T}^{-1} = a_{-\bm k}^{\dagger}, \quad \mathcal{T} b_{\bm k}^{\dagger} \mathcal{T}^{-1} = b_{-\bm k}^{\dagger}.
\end{equation}
Complex conjugation by $\mathcal{T}$ inverts the $y$ component of pseudospins:
\begin{equation}
(n_{\bm k}^x, n_{\bm k}^y, n_{\bm k}^z) \xrightarrow{\mathcal{T}} (n_{-\bm k}^x, -n_{-\bm k}^y, n_{-\bm k}^z).
\end{equation}
If TRS is intact, we have
\begin{equation}
h_{-\bm k}^x = h_{\bm k}^x, \quad h_{-\bm k}^y = -h_{\bm k}^y, \quad h_{-\bm k}^z = h_{\bm k}^z.
\end{equation}
Combined $C_2 \mathcal{T}$ operation inverts only the $z$-component of pseudospins while keeping the momentum labels unchanged:
\begin{equation}
(n_{\bm k}^x, n_{\bm k}^y, n_{\bm k}^z) \xrightarrow{C_2 \mathcal{T}} (n_{\bm k}^x, n_{\bm k}^y, -n_{\bm k}^z).
\end{equation}
When $C_2 \mathcal{T}$ symmetry is preserved, all first-order-in-$n^z$ terms must vanish:
\begin{equation}
    h_{\bm k}^z = J_{\bm k \bm k'}^{zx} = J_{\bm k \bm k'}^{zy} = 0.
\end{equation}

\section{Construction of sublattice basis states} \label{sec:wannierize}

To construct a pair of orthonormal sublattice basis states, we solve the Hamiltonian $H = H_{\rm kin} + H_{\rm SL}$ in which the superlattice potential is honeycomb-shaped ($U_{\bm g} \in \mathbb{R}$) and the form factors are trivial ($\Lambda_{\bm p', \bm p} = 1$). We keep only the first harmonics of the honeycomb potential:
\begin{equation}
U(\bm r) = 2U_1 \sum_{i=1}^3 \cos(\bm g_i \cdot \bm r),
\end{equation}
where $\bm g_1 = (4\pi/3a,0)$ and $\bm g_2, \bm g_3$ are its $C_3$-partners. Expanding the potential around its minimum at $\bm r_0 = (a/\sqrt{3},0)$, at quadratic order we get
\begin{equation}
U(\bm r) \approx -3U_1 + \frac{4\pi^2 U_1}{a^2} (\bm r - \bm r_0)^2.
\end{equation}
Comparison with the harmonic oscillator problem gives the localization length $l = (\hbar^2 a^2 / 8\pi^2 mU_1)^{1/4}$.

In a honeycomb-lattice potential, the lowest two bands touch at the Dirac points and are isolated from higher bands. Wannierization of the lowest two bands results in two orthonormal sublattice basis states localized at the two degenerate potential minima. Although a standard procedure to obtain maximally localized Wannier orbitals exists in the literature \cite{marzari1997maximally, souza2001maximally, marzari2012maximally}, we use a simpler method that yields reasonably localized orbitals.

\begin{figure}
    \centering
    \includegraphics[width=0.6\linewidth]{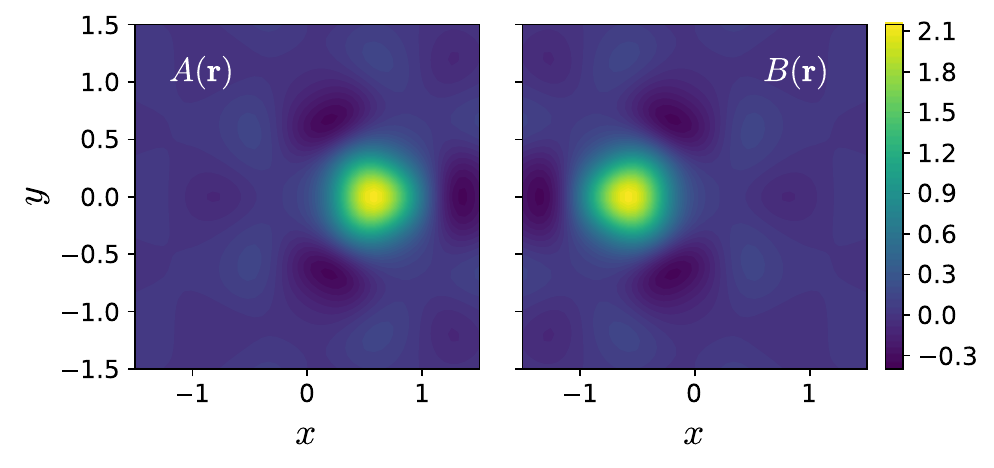}
    \caption{Real-space wavefunctions of sublattice basis states with localization length $l=0.25$. All lengths are expressed in units of lattice constant $a$.}
    \label{fig:wannier_basis}
\end{figure}

When the honeycomb potential is strong, the transformation from the sublattice basis states to the lowest two band states is approximately given by the tight-binding model with only nearest-neighbor hopping. Taking the sublattice basis states located at $\bm a_0 = -\bm b_0 = (a/\sqrt{3},0)$, solution of the tight-binding model gives the lower and upper eigenstates
\begin{align}
\alpha_{\bm k}^{\dagger} &= \frac{1}{\sqrt{2}} (e^{i\eta_{\bm k}/2} a_{\bm k}^{\dagger} + e^{-i\eta_{\bm k}/2} b_{\bm k}^{\dagger}), \\
\beta_{\bm k}^{\dagger} &= \frac{i}{\sqrt{2}} (e^{i\eta_{\bm k}/2} a_{\bm k}^{\dagger} - e^{-i\eta_{\bm k}/2} b_{\bm k}^{\dagger}),
\end{align}
where $\eta_{\bm k} = \arg \left[ 2e^{i\sqrt{3}k_x a/2} \cos(k_y a/2) + e^{i\sqrt{3} k_y a} \right]$. Equivalently, in the plane-wave basis $\alpha_{\bm k}^{\dagger} = \sum_{\bm g} C_{\bm k + \bm g} c_{\bm k + \bm g}^{\dagger}$, $\beta_{\bm k}^{\dagger} = \sum_{\bm g} D_{\bm k + \bm g} c_{\bm k + \bm g}^{\dagger}$, $C$ and $D$ have the expressions
\begin{align}
C_{\bm k + \bm g} &= \frac{1}{\sqrt{2}} (e^{i\eta_{\bm k}/2} A_{\bm k + \bm g} + e^{-i\eta_{\bm k}/2} B_{\bm k + \bm g}), \\
D_{\bm k + \bm g} &= \frac{i}{\sqrt{2}} (e^{i\eta_{\bm k}/2} A_{\bm k + \bm g} - e^{-i\eta_{\bm k}/2} B_{\bm k + \bm g}).
\end{align}
In practice, $\{C_{\bm k + \bm g}\}$ and $\{D_{\bm k + \bm g}\}$ are the lowest two eigenvectors obtained by numerically solving the Hamiltonian in the plane wave basis, aside from an arbitrary phase. To fix the phase, we notice that because $A_{\bm p} = \varphi_{\bm p} e^{-i\bm p \cdot \bm a_0}$ and $B_{\bm p} = \varphi_{-\bm p} e^{-i\bm p \cdot \bm b_0}$ are complex conjugate, $C$ and $D$ in the above expressions are purely real. To fix the sign, notice that because of the approximate circular symmetry of the localized orbitals, $\varphi_{\bm p}$ is approximately real at small $|\bm p|a \lesssim 1$. The signs of $C_{\bm k + \bm g}$ and $D_{\bm k + \bm g}$ are fixed by requiring that the largest component of the numerically obtained eigenvector has the same sign as $\cos[(\bm k + \bm g) \cdot \bm a_0 - \eta_{\bm k}]$ and $\sin[(\bm k + \bm g) \cdot \bm a_0 - \eta_{\bm k}]$, respectively. $A$ and $B$ are the obtained by inverting the above equations:
\begin{align}
A_{\bm k + \bm g} &= \frac{1}{\sqrt{2}} (C_{\bm k + \bm g} - iD_{\bm k + \bm g}) e^{-i\eta_{\bm k}/2}, \\
B_{\bm k + \bm g} &= \frac{1}{\sqrt{2}} (C_{\bm k + \bm g} + iD_{\bm k + \bm g}) e^{i\eta_{\bm k}/2}.
\end{align}
Fig.~\ref{fig:wannier_basis} shows the real-space wavefunctions $A(\bm r) = \sum_{\bm p} A_{\bm p} e^{i\bm p \cdot \bm r} / \sqrt{N_{\rm cell} \mathcal{A}}$ and $B(\bm r) = \sum_{\bm p} B_{\bm p} e^{i\bm p \cdot \bm r} / \sqrt{N_{\rm cell} \mathcal{A}}$ of the sublattice basis states with localization length $l = 0.25 a$. Here $N_{\rm cell}$ is the number of unit cells and $\mathcal{A} = \sqrt{3} N_{\rm cell} a^2/2$ is the total area of the 2D system.

\section{Different winding numbers}

In the numerical calculations in the main text, the Bloch wavefunctions are taken to be winding spinors $\ket{u_{\bm p}} = (\cos(\alpha_{\bm p}/2), e^{i\beta_{\bm p}} \sin(\alpha_{\bm p}/2))$ where $\alpha_{\bm p} = \arctan(\gamma |\bm p|a)$ and $\beta_{\bm p} = N \arg(p_x+ip_y)$. The total Berry flux of the band is $\int d^2 \bm p \, \Omega_{\bm p} = N\pi$. While the Berry curvature clearly increases with winding number $N$, in this section we show that for fixed $\gamma$, the tendency to form AHCs is not monotonic with increasing $N$. Therefore, the Berry flux within the first mBZ is not the only relevant quantity.

\begin{figure}
    \centering
    \includegraphics[width=0.32\linewidth]{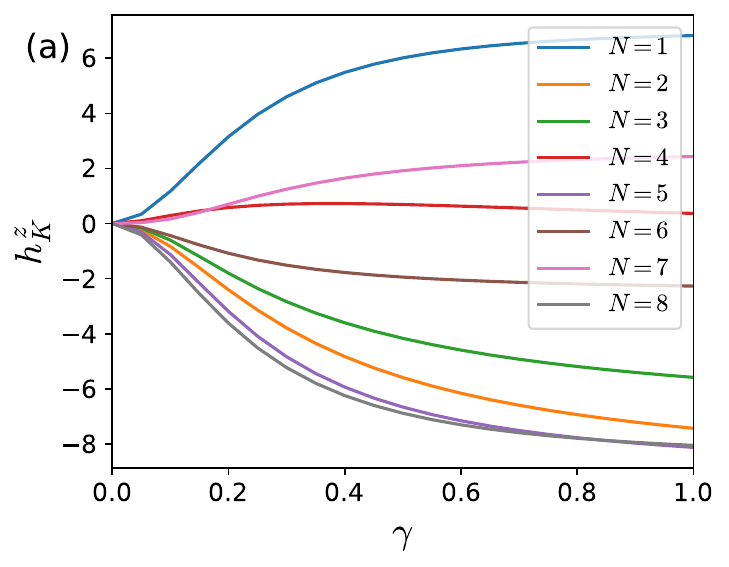}
    \includegraphics[width=0.32\linewidth]{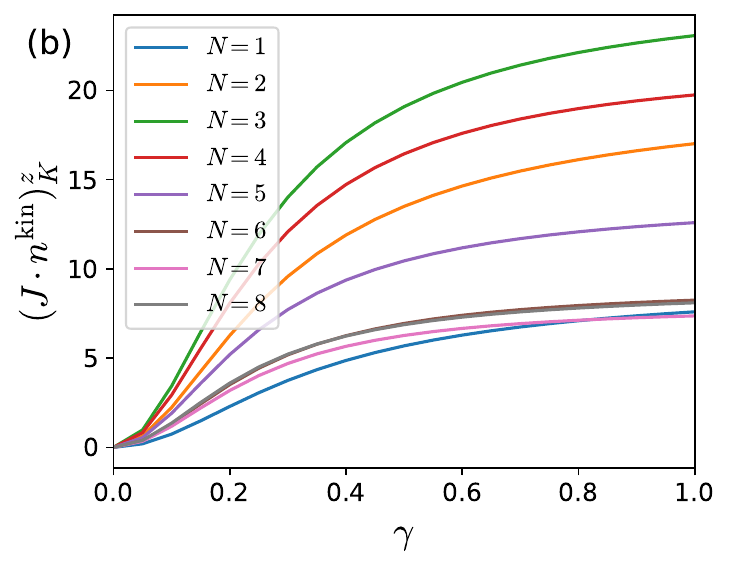}
    \includegraphics[width=0.32\linewidth]{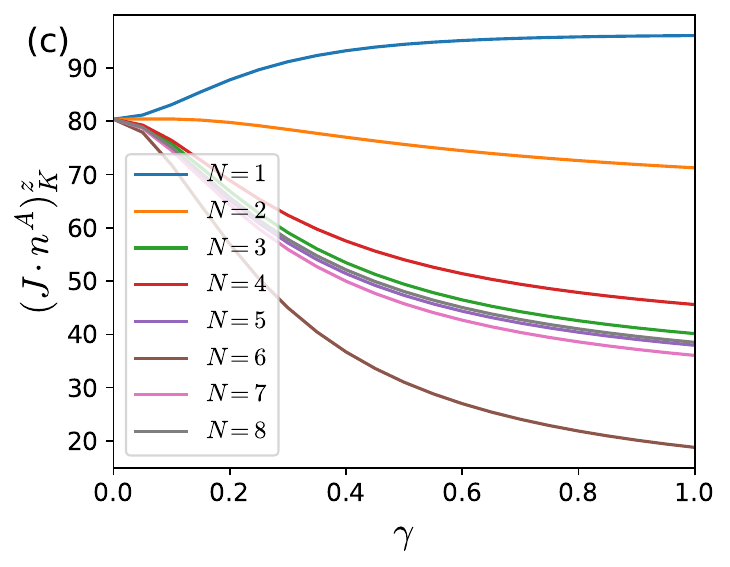}
    \caption{Out-of-plane components of effective pseudospin Zeeman fields at $K$ point as functions of $\gamma$ at $r_s=20$. Curves of different colors in each plot represent results at different winding numbers $N$. Three subfigures respectively represent (a) the pseudospin Zeeman field $h^z$ (linear coefficient of $n^z$ in $E_{\rm MF}$); (b) effective Zeeman field generated by $\bm n^{\rm kin}$ and $J^{zx}, J^{zy}$ couplings; (c) effective Zeeman field generated by the $A$-sublattice polarized state $\bm n^A = (0,0,1)$ and $J^{zz}$ coupling.}
    \label{fig:hz_gam_N}
\end{figure}

As a measure of tendency to form AHC states, in Fig.~\ref{fig:hz_gam_N} we plot the pseudospin Zeeman field $h^z$ and the effective Zeeman fields $J\cdot n^{\rm kin}$ and $J\cdot n^A$ at $K$ as functions of $\gamma$ at different winding numbers $N$. Roughly speaking, AHCs are stabilized when the net sum of the first two quantities is large, and WCs are destabilized when the third quantity is small. It is clear from the numerical results in Fig.~\ref{fig:hz_gam_N} that none of these quantities is monotonic with $N$.

\end{document}